\documentclass{aa}  

\usepackage{graphicx}
\usepackage[dvipsnames]{xcolor}
\usepackage{caption}
\usepackage{multirow}
\usepackage{txfonts}
\usepackage[]{hyperref}

\newcommand\figsize{0.49}
\newcommand\figsizee{0.99}

\newcommand{\xmm}{{\it XMM-Newton}\xspace}

\newcommand{\chandra}{{\it Chandra}\xspace}
\newcommand{\suzaku}{{\it Suzaku}\xspace}

\newcommand{\gaia}{{\it Gaia}\xspace}

\hypersetup{colorlinks,citecolor=blue}
\usepackage{threeparttable}

\usepackage{txfonts}
\begin{document}

   \title{A sample of ionised Fe line-emitting X-ray sources in the inner Galactic disc}


   \author{Samaresh Mondal\inst{1,2}, 
            Gabriele Ponti\inst{1,3}, 
            Tong Bao\inst{1},
            Mark R. Morris\inst{4},
            Frank Haberl\inst{3},
            Nanda Rea\inst{5,6}, and
            Sergio Campana\inst{1}
          }

   \institute{$^1$INAF – Osservatorio Astronomico di Brera, Via E. Bianchi 46, 23807 Merate (LC), Italy
            \email{samaresh.astro@gmail.com}\\
            $^2$Department of Astronomy, University of Illinois, 1002 W. Green St., Urbana, IL 61801, USA\\
            $^3$Max-Planck-Institut für extraterrestrische Physik, Gießenbachstraße 1, 85748, Garching, Germany\\
            $^4$Department of Physics and Astronomy, University of California, Los Angeles, CA, 90095-1547, USA\\
            $^5$Institute of Space Sciences (ICE, CSIC), Campus UAB, Carrer de Can Magrans s/n, E-08193 Barcelona, Spain\\
            $^6$Institut d'Estudis Espacials de Catalunya (IEEC), Carrer Gran Capit\`a 2--4, E-08034 Barcelona, Spain
            }

   \date{Received XXX; accepted YYY}
   \authorrunning{Mondal et al.}
   \titlerunning{a sample of {Fe\,\textsc{xxv}} line-emitting sources in the inner Galactic disc}

 
  \abstract
   {The origin of the unresolved X-ray emission towards the Galactic Centre and the Galactic disc is unclear. Previous studies suggest that the Galactic diffuse X-ray emission is composed of unresolved point sources, primarily magnetic cataclysmic variables (mCVs). However, mCVs have a much lower 6.7 keV line equivalent width ($\rm EW_{6.7}$) and a much higher line intensity ratio between {Fe\,\textsc{xxvi}} and {Fe\,\textsc{xxv}} ($\frac{I_{6.9}}{I_{6.7}}$) compared to the diffuse X-ray emission. Therefore, the primary contributors to the unresolved X-ray emission remain unclear.}
   {We performed a deep X-ray scan of the inner Galactic disc using \xmm observations covering $350^\circ<l<7^\circ$ and $-0.8^\circ<b<+0.8^\circ$. We aim to characterise the different populations of point sources that contribute to the Galactic diffuse X-ray emission by estimating the X-ray spectral slope $\Gamma$ in the 0.2--10 keV band, $\rm EW_{6.7}$, and $\frac{I_{6.9}}{I_{6.7}}$.}
   {We performed source detection in the 6.5--7 keV band. We then extracted the spectra of the X-ray point sources and performed spectral modelling using {\sc xspec} to estimate the X-ray spectral index $\Gamma$, equivalent width, and intensities of the iron 6.7 and 6.9 keV lines.}
   {We detected a total of 859 sources in the 6.5--7 keV band, of which 72 sources show significant iron line emission at 6.7 keV. The distribution of spectral index $\Gamma$ for these 72 sources is bimodal, with peaks at $\Gamma=0.5\pm0.4$ and $1.8\pm0.3$, suggesting two populations of sources. The soft X-ray sources ($\Gamma\sim1.8$) have significantly larger $\rm EW_{6.7}$ than the hard X-ray sources ($\Gamma\sim0.5$). Furthermore, 18 of the 32 hard sources are associated with previously known CVs. We identify CV candidates in our sample as those with spectral index $\Gamma<1.25$. The line ratio, 2--10 keV luminosity, and previous detection of spin period suggest that most of these CVs are magnetic. The distribution of the $\rm EW_{6.7}$ line for the combined sample of previously identified and candidate CVs has a mean value of <$\rm EW_{6.7}$>$=415\pm39$ eV. Furthermore, we computed the stacked spectra of all sources detected in the 6.5--7 keV band for different flux groups, and we find evidence in the stacked spectra of hard sources that the $\rm EW_{6.7}$ increases with decreasing flux. The soft X-ray sources have <$\rm EW_{6.7}$>$=1.1\pm0.1$ keV. We identified 13 of the 30 soft sources associated with active stars, young stellar objects, and active binaries of RS CVn type.}
   {The {Fe\,\textsc{xxv}} line-emitting sources towards the Galactic disc can be broadly categorised into two types: soft and hard X-ray sources. The <$\rm EW_{6.7}$> of our CV candidate sample is more than twice as large as the typical $\rm EW_{6.7}$ found in mCVs within 500 pc, and the <$\rm EW_{6.7}$> of our CV candidate sample is close to the $\rm EW_{6.7}$ value of Galactic diffuse X-ray emission. In our survey, the detection sensitivity for point sources in the 6.5--7 keV band is $\sim10^{-14}\ \rm erg\ s^{-1}\ cm^{-2}$. Therefore, up to a flux limit of $\sim 10^ {-14}\ \rm erg\ s^{-1}\ cm^{-2}$ or higher, nearly 50\% of Galactic diffuse X-ray emission in the 6.5--7 keV band originates from mCVs. The large <$\rm EW_{6.7}$> of the soft source sample indicates that these sources also contribute significantly to the Galactic diffuse X-ray emission, as well as from CVs.}

   \keywords{Galaxy: bulge -- Galaxy:centre -- Galaxy:disc -- novae, cataclysmic variables -- white dwarfs -- X-rays: binaries}

   \maketitle
   
%
\section{Introduction}
Early X-ray observations have revealed a large-scale, diffuse X-ray emission of Galactic origin \citep[see][for a review of X-ray activity in the Galactic Centre]{ponti2013}. The Galactic diffuse X-ray emission has a surface brightness slightly higher than that of the cosmic X-ray background, but it is more concentrated along the Galactic plane, with a scale height of only a few hundred parsecs \citep{yamauchi2016}. This emission can be decomposed into three spatial components: Galactic centre X-ray emission (GCXE; \citealt{kellogg1971,watson1981}), Galactic bulge X-ray emission (GBXE; \citealt{cooke1969,warwick1980}), and Galactic ridge X-ray emission (GRXE; \citealt{worrall1982,warwick1985}). The Galactic diffuse X-ray emission shows multiple iron emission lines, with the K-shell transition line from He-like Fe ({Fe\,\textsc{xxv}}-He$\alpha$) at 6.7 keV being the brightest. The 6.7 keV iron line emission is a significant feature in X-ray astronomy, particularly in the context of our Galaxy \citep{koyama2018}. This emission is primarily associated with ionised iron and provides critical insights into the physical conditions and processes occurring in the Galactic environment. \citet{uchiyama2013} performed a survey with \suzaku of the Galactic Centre (GC) region and additional regions at higher longitude $l$ and higher latitude $b$, demonstrating that the intensity of the {Fe\,\textsc{xxv}} line increases towards the GC. The overall Galactic diffuse X-ray emission spectrum can be fitted by a two-temperature, collisionally ionised plasma model with temperatures of $\sim0.75$ and $\sim7.5$ keV \citep{nobukawa2016}.

The 6.7 keV line emission serves as a diagnostic tool for studying hot plasma environments, which are indicative of energetic processes such as supernova remnants and active stellar populations \citep{porquet2010}. Initially, the Galactic diffuse X-ray emission was interpreted as hot plasma confined within the Galactic potential well \citep{watson1981,warwick1985}. However, the Galactic potential well is too shallow to confine such a hot plasma \citep{tanaka2002,belmont2005}. The correlation between the 6.7 keV line intensity of Galactic diffuse X-ray emission and the near-infrared surface brightness across different regions supports a link to star formation and stellar activity \citep{revnivtsev2006}. A recent study by \citet{anastasopoulou2023} showed that the diffuse hard emission in the central degrees at the GC can be explained by assuming a GC stellar population with iron abundances $\sim$1.9 times higher than those in the Galactic bar or bulge. In some regions of the Galactic diffuse X-ray emission, \chandra has resolved 80\% of the 6.7 keV emission as X-ray point sources \citep{revnivtsev2009}. The hotter $\sim7.5$ keV emission of the Galactic diffuse X-ray emission is thought to originate from unresolved, accreting magnetic cataclysmic variables (mCVs). A similar study by \citet{hong2012} suggested that, above 3 keV, 80\% of the resolved sources are mCVs, while the net emission from active binaries (ABs) constitutes the remaining 20\%. A recent study by \citet{koyama2024} attempted to fit the spectra of the GCXE, GRXE, and GBXE with the composite spectra of all local white dwarfs (WD) and X-ray active stars, suggesting that additional components are still required, which may indicate the presence of truly diffuse emission originating from supernova remnants.

The emission from mCVs, including intermediate polars (IPs) and polars, is dominated by hard X-ray emission and ionised lines between 6.5--7.0 keV. In mCVs, accretion shocks above the WD surface heat the infalling material, leading to temperatures of $kT>15$ keV. These extremely high temperatures ionise the medium, which emits the {Fe\,\textsc{xxv}} and {Fe\,\textsc{xxvi}} lines at 6.7 and 6.9 keV, respectively. In addition, mCVs exhibit fluorescent Fe $K_{\alpha}$ line emission at 6.4 keV, originating from the reflection of X-rays off the WD surface \citep[see][for a review on accreting WD binaries]{cropper1990,petterson1994,mukai2017}. \citet{ezuka1999} studied a sample of 20 magnetic CVs and found average equivalent widths (EWs) of  $\rm EW_{6.4}\sim100$, $\rm EW_{6.7}\sim170$, and $\rm EW_{6.9}\sim100$ eV, for the  6.4, 6.7, and 6.9 keV lines, respectively. Similar EW values were reported by \citet{hellier1998}. Furthermore, \citet{hellier2004} investigated a sample of five mCVs using \chandra high energy transmission grating spectra and found average EW values of $\rm EW_{6.4}\sim120$, $\rm EW_{6.7}\sim160$, and $\rm EW_{6.9}\sim110$ eV. More recently, \citet{xu2016} studied a sample of 20 mCVs and found mean $\rm EW_{6.7}$ values of $107\pm16$ eV for IPs and $221\pm135$ for polars. Previous studies have highlighted that the EWs of the 6.4 and 6.9 keV iron lines in the mCV population are similar to those in the Galactic diffuse X-ray emission. However, their $\rm EW_{6.7}$ values are 2$-$3 times smaller than in the GCDE or GRXE. 

The line intensity ratio between {Fe\,\textsc{xxvi}} and {Fe\,\textsc{xxv}} ($\frac{I_{6.9}}{I_{6.7}}$) in the mCV population also differs significantly from the Galactic diffuse X-ray emission. For example, the GCXE and GRXE have $\rm EW_{6.7}$ values of $500\pm3$ and $487\pm13$ eV, respectively \citep{nobukawa2016}. The line intensity ratio $\frac{I_{6.9}}{I_{6.7}}$ for the GCXE and GRXE is $0.367\pm0.005$ and $0.16\pm0.02$, respectively \citep{koyama2018}. In contrast, the mean $\rm EW_{6.7}$ of the mCV population in the solar neighbourhood is  $\sim160$ eV \citep{ezuka1999,hellier1998,xu2016}, and the intensity ratio $\frac{I_{6.9}}{I_{6.7}}$ of mCVs in the solar neighbourhood varies between $0.4-0.7$ \citep{xu2016}. For the active binary (AB) population, both the $\rm EW_{6.7}$ and the $\frac{I_{6.9}}{I_{6.7}}$ values are significantly lower than those of the GCXE and GRXE. The AB population typically has an $\rm EW_{6.7}$ of $\sim327$ eV and a line intensity ratio of $\sim0.14$ \citep{nobukawa2016}. As the $\rm EW_{6.7}$ and $\frac{I_{6.9}}{I_{6.7}}$ values of mCV and AB populations do not match those of the GCXE and GRXE, it remains unclear how the GCXE and GRXE decompose into specific X-ray source classes and what their contributions are to the total emission component. In this study, we use the $\rm EW_{6.7}$ of the ionised {Fe\,\textsc{xxv}} line and the line intensity ratio $\frac{I_{6.9}}{I_{6.7}}$ as diagnostic tools to probe the temperature and metallicity of the X-ray emitting plasma, and to constrain the nature and population of the underlying source classes responsible.

\section{Data reduction}
\label{data_reduction}
We conducted an X-ray scan (PI: G. Ponti) of the inner Galactic disc covering $350^\circ<l<7^\circ$ and $-0.8^\circ<b<+0.8^\circ$. Details of the Heritage survey, including the multi-band X-ray mosaic of the region, will be presented in a forthcoming paper (Ponti et al., in prep). We analysed a total of 588 \xmm \citep{jansen2001} observations, including the Heritage data for the inner Galactic disc and archival observations of the Galactic Centre \citep{ponti2015,ponti2019}. The observation data files were processed using the \xmm Science Analysis System (SASv19.0.0)\footnote{https://www.cosmos.esa.int/web/xmm-newton/sas}. This work is largely based on the \xmm data set presented in \citet{mondal2024a}, with slightly different selection criteria, as described below. High background flaring activity was filtered out after careful inspection. We removed the time interval of increased flaring activity when the background was above a threshold of 8 counts ks$^{-1}$ arcmin$^{-2}$ for EPIC-pn \citep{struder2001} and 2.5 counts ks$^{-1}$ arcmin$^{-2}$ for EPIC-MOS1/MOS2 \citep{turner2001} in the 7--15 keV bands. We used the SAS task \texttt{emldetect} to detect point sources and generate the source list. Detection was performed in five energy bands: 1--4.5, 4.5--6, 6--6.5, 6.5--7, and 7--12 keV. We computed the maximum likelihood (ML) and net counts in the broadband 1--12 keV band from the sub-bands. We chose \texttt{EXT==0} to select only the point sources. We also applied an ML cut to detect sources that were more likely to be real. Therefore, to construct the cleanest sample, we chose $\rm ML>14$, where all sources are statistically real with a $>5\sigma$ confidence level \citep{webb2020}. We detected a total of 859 sources with $\rm ML>14$ in the 6.5--7 keV band. We extracted the spectra using the SAS task \texttt{evselect}. We selected events with PATTERN$\le$4 and PATTERN$\le$12 for EPIC-pn and MOS1/MOS2 detectors, respectively. We chose a circular region of 20\arcsec\ radius for the source product extraction. The background products were extracted from an annular region centred on the source position, with inner and outer radii of 25\arcsec\ and 30\arcsec, respectively. The spectra from each detector (pn, MOS1, or MOS2) were grouped to have a signal-to-noise ratio (S/N) of 3 in each energy bin.

\begin{figure*}
\centering
\includegraphics[width=\figsizee\textwidth]{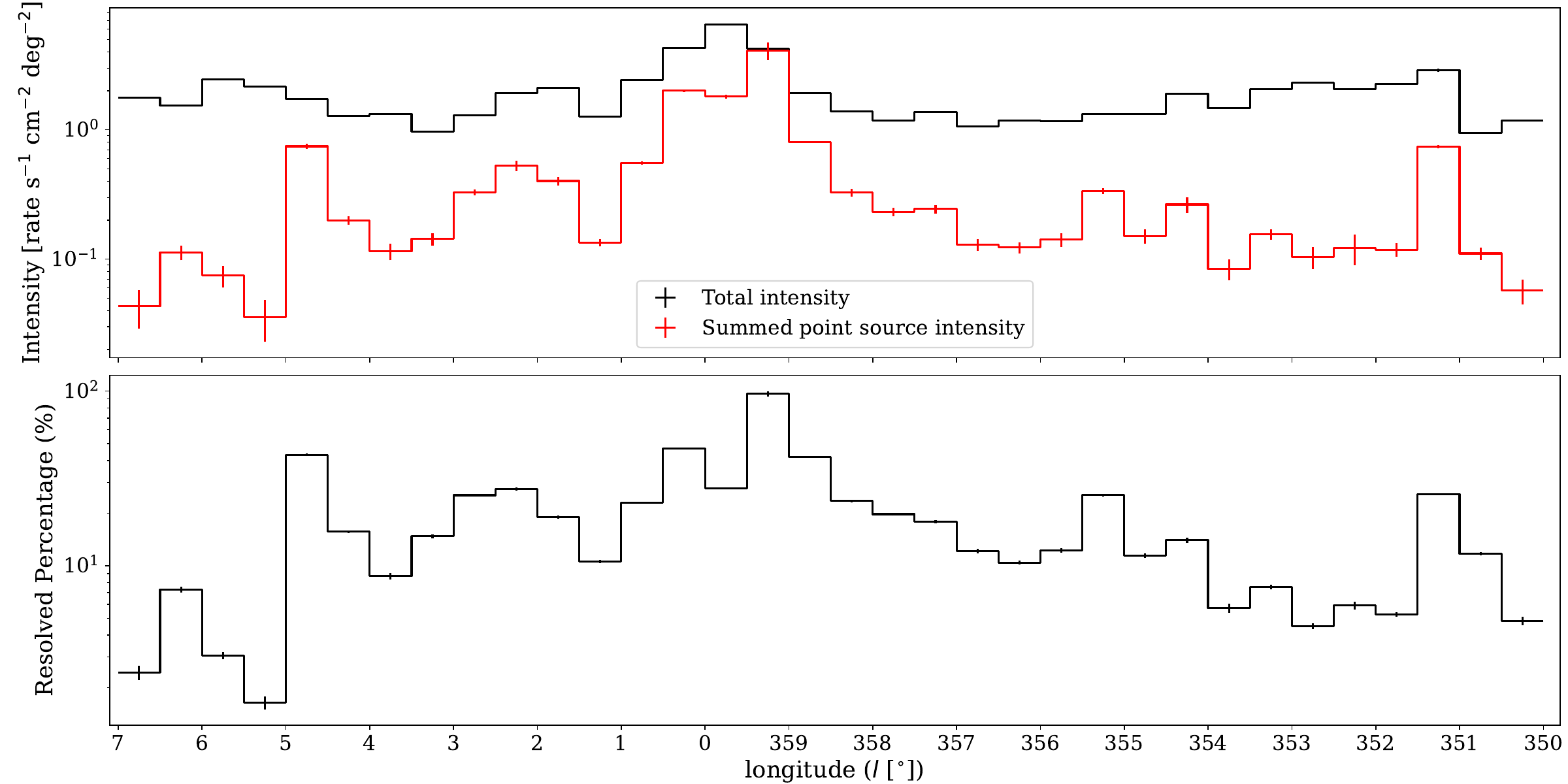}
\caption{Top panel: Longitude profile of total emission (point sources plus unresolved emission) in the 6.5--7.0 keV range, integrated over $-0.8^{\circ}\leq b\leq0.8^{\circ}$ (black curve). The red curve shows the contribution from point sources detected with $\rm ML>14$ in the 1--12 keV band. Bottom panel: Percentage of the total 6.5--7.0 keV emission resolved by X-ray point sources.}
\label{fig:emission_profile}
\end{figure*}
\section{Spectral fitting procedure}
\label{sec:spec_fit}
We utilised PyXspec, a Python interface to {\sc xspec} for automated X-ray spectral fitting \citep{arnaud1996,gordon2021}. We used $\chi^2$ statistics in our spectral modelling and estimated the 95\% error for each fitting parameter. Before estimating the iron line EW, it is essential to select the correct continuum model that properly fits the 0.2--10 keV X-ray spectra. The choice of the X-ray continuum model is crucial in estimating the significance and EW of the detected line. We tested two continuum models: a power-law model and a blackbody plus power-law model. If the continuum model does not adequately describe the data, the estimation of line intensity and EW can be significantly biased. In sources whose emission is dominated by soft photons, their continuum might not be well described by a simple absorbed power-law model, potentially leading to errors in EW estimation. Therefore, for each source, we compared the fit quality of the simple power-law and blackbody plus power-law models by examining their $\chi^2$ and degrees of freedom ($d.o.f$). We favoured the blackbody plus power-law continuum model over the simple power-law model if it improved the continuum fitting above the $3\sigma$ confidence level. During the fit, we included an absorption component \texttt{tbabs} with the default abundance value \texttt{wilm} \citep{wilms2000} to account for the absorption of X-rays along the line of sight by the interstellar medium.

For each source, we first fitted the spectrum with the selected continuum model and recorded the fit parameters, $\chi^2$ and $d.o.f$ values. We then added a Gaussian component at 6.7 keV and computed the fit statistics. We fixed the centroid of the line at 6.7 keV and the line width at zero eV while performing the fit. The \xmm spectral resolution cannot resolve the thermal broadening of the line; therefore, we fixed the width of the line at zero eV. We evaluated the improvement in fit by calculating $\Delta\chi^2$ and computed the statistical significance of the line using an F test. Next, we computed the EW and the intensity of the line. We repeated this procedure for the {Fe\,\textsc{xxvi}} line by adding a Gaussian at 6.9 keV and similarly computed the EW, line intensity, and significance of the line.

\section{Results}
\subsection{Emission profile and resolved percentage}
We compared the total X-ray emission in the 6.5--7 keV band with the emission from resolved point sources. To compute the emission profile, we selected a strip of $-0.8^{\circ}\leq b\leq+0.8^{\circ}$ and longitude $l$ that extended from $350^{\circ}$ to $7^{\circ}$. We constructed a count-rate mosaic map from all \xmm observations considered in this study, which was subsequently smoothed using a Gaussian kernel of $\sigma=5$ pixels; each pixel has a size of $4\times4$ arcsec$^{2}$. We computed the intensity profile considering a sliding box with a longitude bin size $\Delta l=0.5^{\circ}$ and $b=\pm0.8^{\circ}$. We then calculated the total intensity of a given sky area in the 6.5--7 keV band (resolved point sources plus unresolved emission). Furthermore, we also computed the intensity of the 6.5--7 keV emission from point sources only. The top panel of Fig. \ref{fig:emission_profile} shows the longitudinal profile of the total emission intensity (black curve). The red curve shows the intensity of the X-ray emission from point sources only. During the calculation of the intensity profile, the count rate was averaged over the pn, MOS1, and MOS2 detectors.

The bottom panel of Fig. \ref{fig:emission_profile} presents the percentage of total emission resolved into X-ray point sources at different longitudes. The resolved percentage varies depending on the longitude. On average, we resolve $18\%$ of the total emission in the 6.5--7 keV band into point sources. Very close to the GC, the resolved percentage increases to $96\%$, primarily because the total emission is dominated by three well-known, extremely bright point sources.

\subsection{Sample properties}
\begin{figure}[]
\centering
\includegraphics[width=\figsize\textwidth]{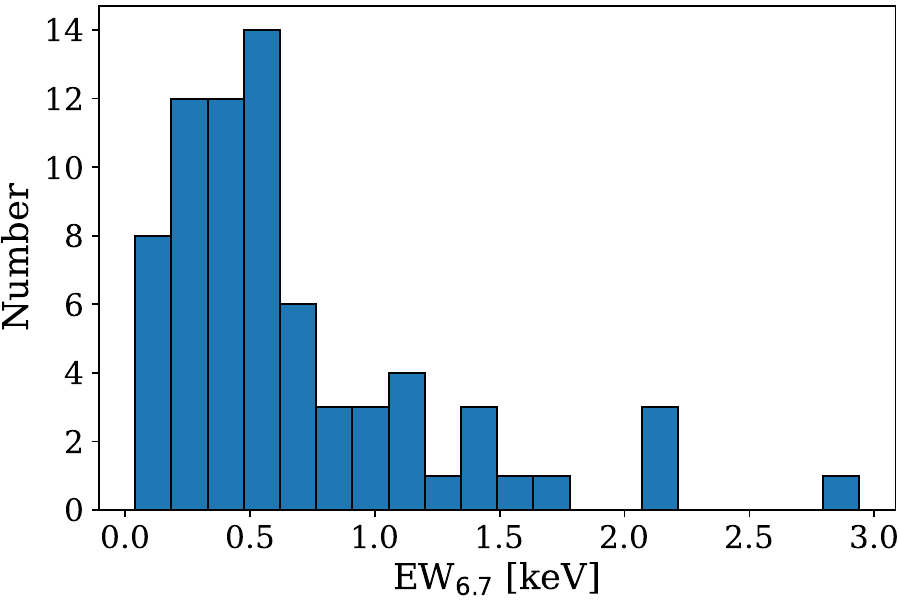}
\includegraphics[width=\figsize\textwidth]{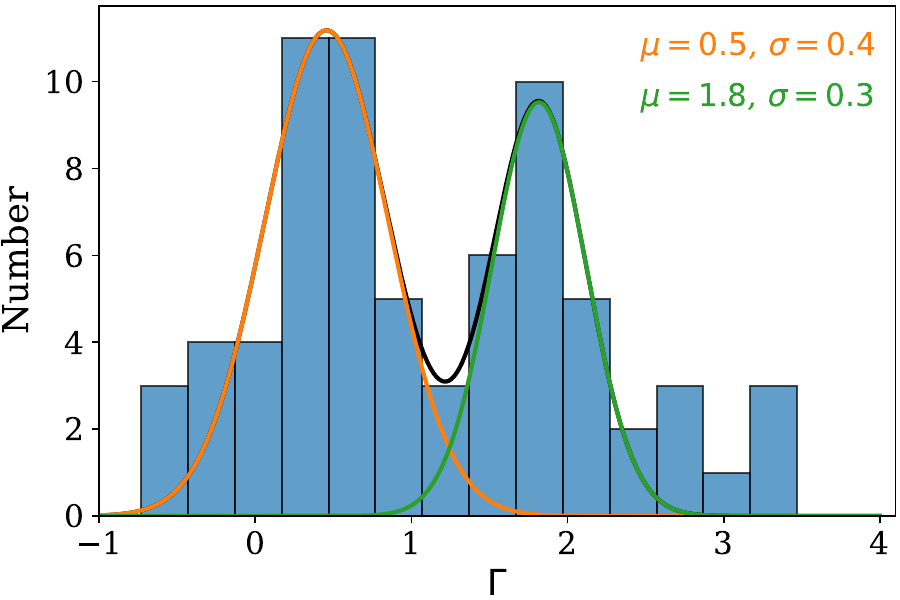}
\caption{Distribution of  EW of the 6.7 keV line (top panel) and X-ray spectral index $\Gamma$ for 6.7 keV line sources (bottom panel).}
\label{fig:ew67_gamma}
\end{figure}

We fit the spectra of all X-ray point sources detected in the 6.5--7 keV band following the method described in Sect.\,\ref{sec:spec_fit}. We selected only sources that showed the presence of a 6.7 keV line above the $2\sigma$ confidence level. We detected the {Fe\,\textsc{xxv}} line above this threshold in a total of 72 sources. The details of these 72 sources are listed in Table \ref{table:list_source}. The top panel of Fig. \ref{fig:ew67_gamma} shows the distribution of $\rm EW_{6.7}$ for these sources. The bottom panel of Fig. \ref{fig:ew67_gamma} shows the distribution of spectral index $\Gamma$, obtained by fitting the spectrum in the 0.2--10 keV band using an absorbed power-law model. The distribution clearly exhibits two peaks at $\Gamma\simeq0.5$ and 1.8, suggesting two distinct source populations. We performed a Kolmogorov–Smirnov (KS) test to determine whether the sample could be drawn from a single normal distribution. The KS test yielded a p-value of $1.5\times10^{-12}$, rejecting the null hypothesis with more than $5\sigma$ confidence. To test the alternative hypothesis of two distinct populations, we fit the distribution with two normal distributions and applied the KS test between the original sample and the distribution drawn from the fitted Gaussians. The p-value for this alternative hypothesis is 0.93, indicating the presence of two populations of sources. This becomes even clearer when comparing the $\Gamma$ and  $\frac{I_{6.9}}{I_{6.7}}$ values of identified sources from SIMBAD, as discussed in Sect.\,\ref{sec:countepart}.

\begin{figure}[]
\centering
\includegraphics[width=\figsize\textwidth]{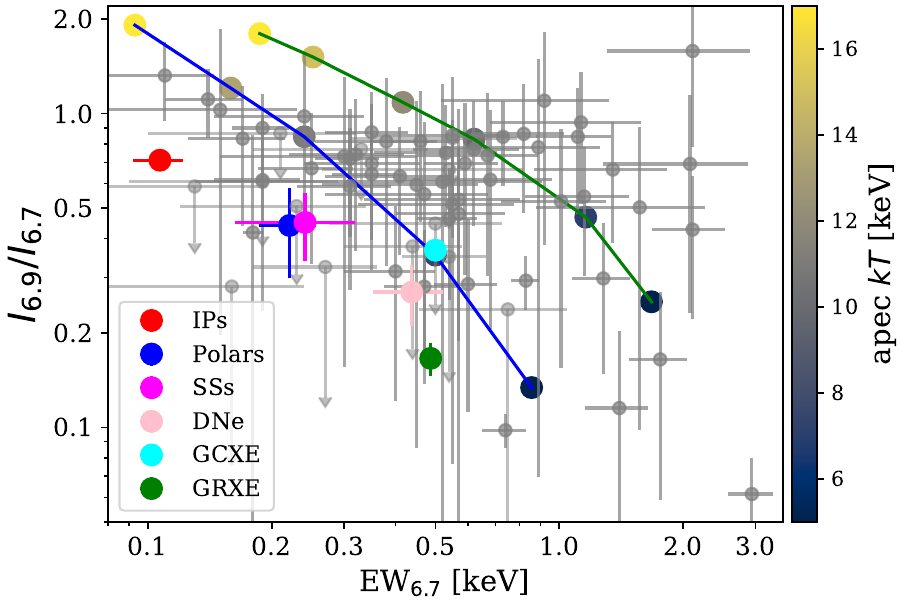}
\caption{Line intensity ratio $\frac{I_{6.9}}{I_{6.7}}$ vs $\rm EW_{6.7}$ plot for sources emitting a 6.7 keV line. Error bars indicate $1\sigma$ uncertainty. Mean values of $\frac{I_{6.9}}{I_{6.7}}$ and $\rm EW_{6.7}$ for different CV-types from \citet{xu2016} are shown for comparison. The $\rm EW_{6.7}$ and $\frac{I_{6.9}}{I_{6.7}}$ values of the GCXE and GRXE are taken from \citet{nobukawa2016}. The dots with blue (abundance $Z_{\odot}$) and green (abundance 2$Z_{\odot}$) lines shows the values of  $\rm EW_{6.7}$ and $\frac{I_{6.9}}{I_{6.7}}$ expected from collisionally ionised plasma at different temperatures.}
\label{fig:ew2_line_ratio}
\end{figure}

\subsubsection{Intensity ratio of ionised Fe lines $\frac{I_{6.9}}{I_{6.7}}$ vs $\rm EW_{6.7}$}
\begin{figure}[]
\centering
\includegraphics[width=\figsize\textwidth]{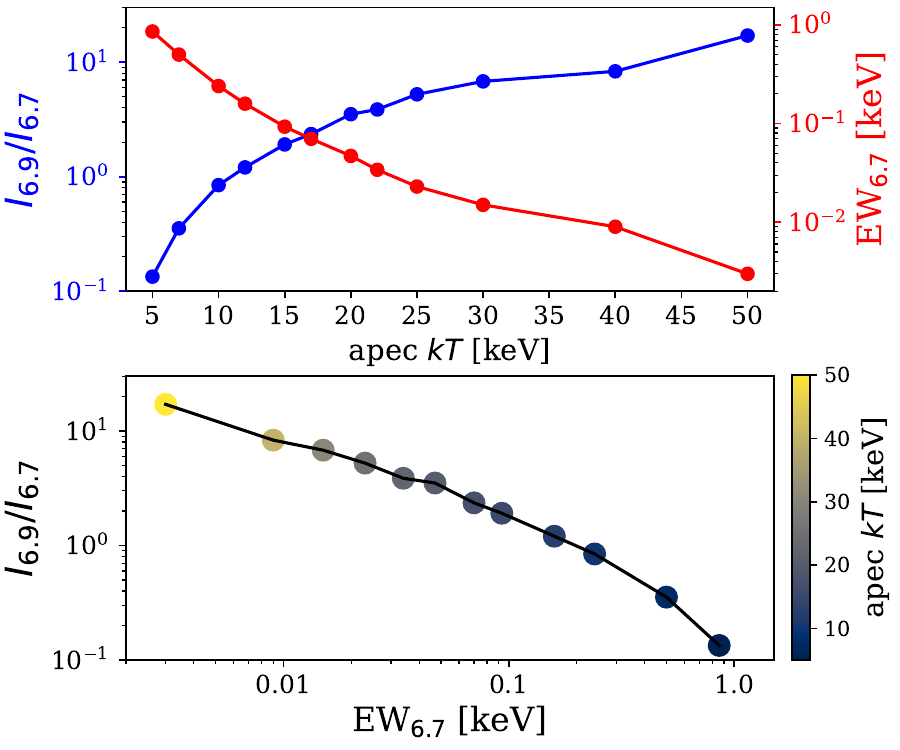}
\caption{Top panel: Relation of $\rm EW_{6.7}$ and $\frac{I_{6.9}}{I_{6.7}}$ with the plasma temperature $kT$ obtained through spectral simulation of ionised plasma of solar abundance. $\rm EW_{6.7}$ and $\frac{I_{6.9}}{I_{6.7}}$ are inversely and positively correlated with $kT$, respectively. Bottom panel: Overall inverse relation between $\rm EW_{6.7}$ and $\frac{I_{6.9}}{I_{6.7}}$.}
\label{fig:kt_ew2_line_ratio}
\end{figure}

We computed the line intensity ratio between {Fe\,\textsc{xxvi}} and {Fe\,\textsc{xxv}} ($\frac{I_{6.9}}{I_{6.7}}$). Among the 72 sources that show {Fe\,\textsc{xxv}} above the $2\sigma$ confidence level, 55 emit a strong {Fe\,\textsc{xxvi}} line. For 13 sources, only upper limit measurements on the {Fe\,\textsc{xxvi}} line were available. For the remaining four sources, the {Fe\,\textsc{xxvi}} line measurement was not possible due to insufficient S/N above 6.9 keV. Figure \ref{fig:ew2_line_ratio} shows the line intensity ratio $\frac{I_{6.9}}{I_{6.7}}$ versus $\rm EW_{6.7}$ for the 68 sources with errors at the $1\sigma$ confidence level. In general, the relationship between $\frac{I_{6.9}}{I_{6.7}}$ and $\rm EW_{6.7}$ exhibits a negative trend. For comparison, we also plot the mean values of $\rm EW_{6.7}$ and line ratio of different mCVs (IPs, polars), non-mCVs (dwarf novae: DNe), and symbiotic systems (SSs) from \citet{xu2016}. The blue and green lines with dots represent the $\rm EW_{6.7}$ and $\frac{I_{6.9}}{I_{6.7}}$ values expected for collisionally ionised plasma at different temperatures. The line intensity ratio is often used to trace the plasma temperature of the material emitting the ionised lines. In general, mCVs like IPs and polars have temperatures much higher than those of non-mCVs (mostly DNe) and ABs. The higher temperature naturally leads to a higher {Fe\,\textsc{xxvi}} ions and a lower fraction of {Fe\,\textsc{xxv}} ions, resulting in an increased value of $\frac{I_{6.9}}{I_{6.7}}$. Therefore, $\frac{I_{6.9}}{I_{6.7}}$ is positively correlated with plasma temperature. The $\rm EW_{6.7}$ also depends on the temperature. Sources with lower temperatures exhibit larger $\rm EW_{6.7}$ values than those with very high plasma temperatures because higher temperatures produce more H-like Fe ions relative to He-like ions. Hence, $\rm EW_{6.7}$ is inversely correlated with temperature. We simulated spectra using {\sc xspec} for a range of plasma temperatures $kT$. The top panel of Fig. \ref{fig:kt_ew2_line_ratio} shows the positive correlation of $\frac{I_{6.9}}{I_{6.7}}$ with temperature $kT$ indicated by the blue curve, and the negative correlation of $\rm EW_{6.7}$ with $kT$ highlighted by the red curve. These correlations result in the overall negative trend between $\frac{I_{6.9}}{I_{6.7}}$ and $\rm EW_{6.7}$ observed in Fig. \ref{fig:ew2_line_ratio}, where the smaller values of $\rm EW_{6.7}$ and the higher value of $\frac{I_{6.9}}{I_{6.7}}$ are associated with sources with high plasma temperatures. In contrast, larger values of $\rm EW_{\rm 6.7}$ and smaller values of $\frac{I_{6.9}}{I_{6.7}}$ are probably associated with lower temperatures. This trend is also evident in the spectral simulation of ionised plasma shown in the bottom panel of Fig. \ref{fig:kt_ew2_line_ratio}.

\begin{figure}[]
\centering
\includegraphics[width=\figsize\textwidth]{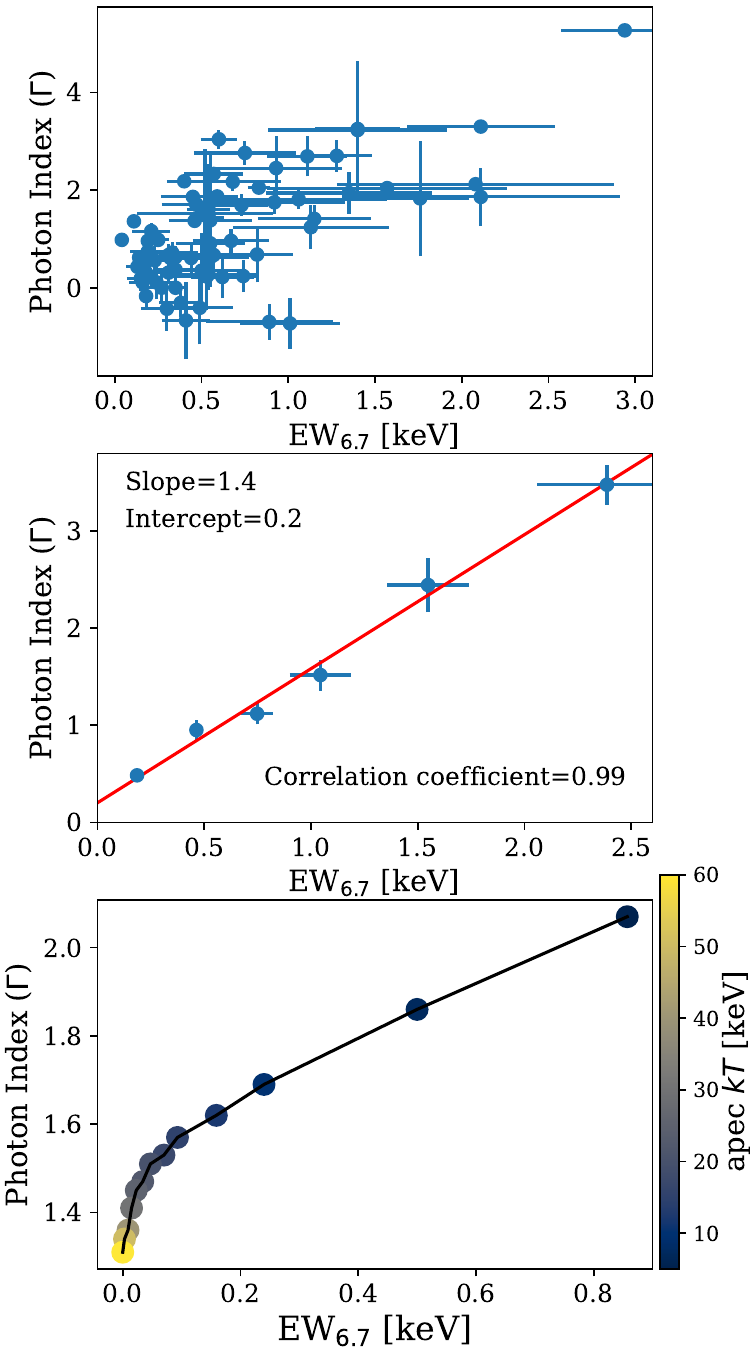}
\caption{Top panel: Scatter plot of photon index $\Gamma$ vs $\rm EW_{6.7}$ of the 72 sources listed in Table \ref{table:list_source}. Middle panel: same as the above plot, with data re-binned with an EW step size of 0.3 keV and errors propagated from individual measurement. Bottom panel: Linear correlation between $\Gamma$ vs $\rm EW_{6.7}$ obtained from spectral simulation of ionised plasma at different temperatures of solar abundance.}
\label{fig:gamma_ew2}
\end{figure}
\subsubsection{Correlation of $\Gamma$ and $\rm EW_{6.7}$}
If the inverse correlation of $\frac{I_{6.9}}{I_{6.7}}$ versus $\rm EW_{6.7}$ is driven by temperature, then some level of correlation between $\rm EW_{6.7}$ and the X-ray spectral slope, $\Gamma$, is expected, as $\Gamma$ is influenced by the temperature of the system. In the upper panel of Fig. \ref{fig:gamma_ew2} we plot $\rm EW_{6.7}$ versus $\Gamma$ for the 72 sources. The scatter plot shows an apparent linear positive correlation, suggesting that soft sources with large values of $\Gamma$ have higher $\rm EW_{6.7}$ than hard sources with smaller $\Gamma$. The correlation is more pronounced when we bin the data with a constant step size of 0.3 keV in $\rm EW_{6.7}$ (middle panel of Fig. \ref{fig:gamma_ew2}). We computed the Pearson correlation coefficient of the grouped data, obtaining a value of 0.99, indicating a strong positive correlation. The correlation has a p-value of 0.0016, indicating that a linear relationship between $\rm EW_{6.7}$ and $\Gamma$ is present above the $3\sigma$ confidence level. Furthermore, fitting the data with a straight line yields a slope of 1.4 and an intercept of 0.2. This linear relationship can be understood by assuming that the X-ray emission from the sources can be modelled with a collisionally ionised plasma model. The bottom panel of Fig. \ref{fig:gamma_ew2} shows a similar linear correlation obtained from spectral simulations of collisionally ionised plasma at different temperatures, where the simulated spectra are fitted with a power-law model. Sources with smaller values of $\Gamma$ represent a higher plasma temperature, resulting in a lower fraction of {Fe\,\textsc{xxv}} atoms and hence lower values of $\rm EW_{6.7}$. For example, the X-ray spectrum of a source with $kT=5$ keV can be modelled by $\Gamma=2.1$ and $\rm EW_{6.7}=0.9$ keV, while a source with $kT=15$ keV will have an X-ray spectrum of $\Gamma=1.5$  and $\rm EW_{6.7}=0.14$ keV. Sources with higher values of $\Gamma$ correspond to very low temperatures, yielding a higher fraction of {Fe\,\textsc{xxv}} atoms and higher $\rm EW_{6.7}$. 

The linear correlation between $\Gamma$ and $\rm EW_{6.7}$ suggests that the X-ray emission from these sources is coming from a thermal mechanism. Furthermore, the bottom panel of Fig. \ref{fig:gamma_ew2} indicates that in our spectral simulation using the \texttt{apec} model, the highest temperature corresponds to a minimum $\Gamma\sim1.3$, whereas the top panel shows many sources with $\Gamma$ below 1. This indicates that a simple model may be insufficient for spectral modelling. Hard sources such as IPs often have local absorption, modelled with a partial covering factor. Furthermore, the physical mechanism that describes the X-ray emission from IPs and polars involves a multi-temperature accretion column, which cannot be accounted for by a single-temperature model such as \texttt{apec} or \texttt{bremsstrahlung}.

\subsection{Counterpart search and classification}
\label{sec:countepart}
\begin{figure}[]
\centering
\includegraphics[width=\figsize\textwidth]{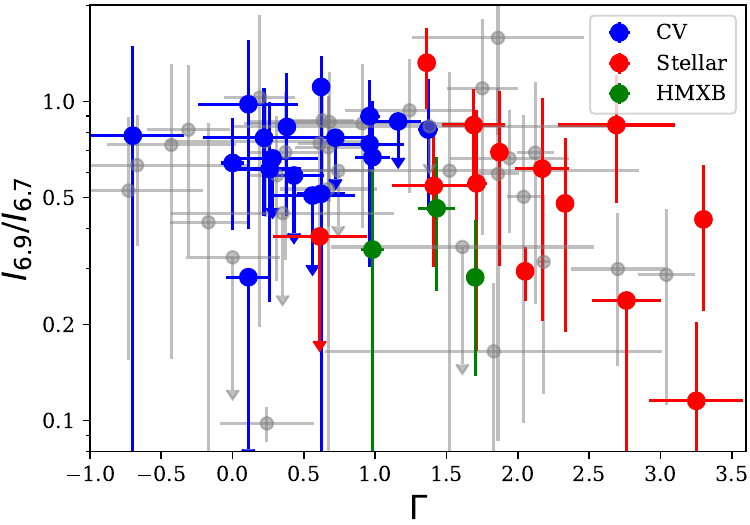}
\caption{Line ratios vs the X-ray spectral photon index $\Gamma$. The small grey points indicate the unidentified sources.}
\label{fig:gamma_line_ratio}
\end{figure}

For the classification of the 72 sources, we searched for counterparts in SIMBAD\footnote{https://simbad.u-strasbg.fr/simbad/} within a 3\arcsec distance from \xmm positions. We found counterparts for almost half of the sources (33 of 72). The counterparts can be broadly categorised into three subcategories: CV, high-mass X-ray binary (HMXB), and stellar association. The majority of the CVs and HMXBs were securely identified in our previous work by detecting their X-ray spin periods \citep{mondal2024a}. In our sample, there are 18 identified CVs which typically have hard X-ray spectra. Of these 18 identified CVs, only two have $\Gamma>1.0$; the rest have X-ray spectral photon indices $\Gamma\leq1.0$. These sources are shown as blue points in Fig. \ref{fig:gamma_line_ratio}.

In the sample of 72 sources, three are classified as HMXBs. All three have X-ray spectral indices $\Gamma\gtrsim1.0$, shown as green points in Fig. \ref{fig:gamma_line_ratio}. The HMXBs are powered by wind accretion, and their strongest emission line in their spectra is the Fe $K_{\alpha}$ emission at 6.4 keV, produced primarily by fluorescence of surrounding neutral material \citep{torrejon2010}. Some HMXBs exhibit highly variable {Fe\,\textsc{xxv}} line emission, while the {Fe\,\textsc{xxvi}} line emission is much weaker compared to {Fe\,\textsc{xxv}} \citep{vrtilek2001,goldstein2004,vanderMeer2005,masetti2010}.

Furthermore, five sources are associated with young stellar objects (YSOs), two associated with RS CVn-type ABs, and six with coronally active stars. All these stellar objects share the common property of having very soft spectra; therefore, we combined them into a single group and designated them as stellar associations. The sources in this category have much softer spectra, with $\Gamma>1.36$, except for one star with $\Gamma=0.61$. These are shown as red points in Fig. \ref{fig:gamma_line_ratio}.

The various source types have different mechanisms for X-ray emission, and among our identified sources, the CVs exhibit a harder spectral shape than the HMXBs and stellar-type sources. Therefore, the higher plasma temperature associated with the harder spectra of CVs leads to a relatively higher $\frac{I_{6.9}}{I_{6.7}}$ ratio of {Fe\,\textsc{xxvi}} emission, relative to the other two types. Figure \ref{fig:gamma_line_ratio} shows the line intensity ratio $\frac{I_{6.9}}{I_{6.7}}$ versus the X-ray spectral index $\Gamma$. It is evident from Fig. \ref{fig:gamma_line_ratio}  that all identified CV-type sources cluster in a restricted region with $\Gamma<1.5$ and $0.3<\frac{I_{6.9}}{I_{6.7}}<1.1$. Among the unidentified sources with $\Gamma<1.5$, only one has $\frac{I_{6.9}}{I_{6.7}}$ below 0.3, indicating that the majority of the hard unidentified sources are likely to be CVs. For the unidentified sources, we classify those with $\Gamma<1.25$ as likely CV candidates. In contrast to the identified CVs, the identified stellar-type sources exhibit a wider range of line ratio values, with $0.1<\frac{I_{6.9}}{I_{6.7}}<1.3$. Furthermore, the stellar-type identified sources show a linear trend of decreasing $\frac{I_{6.9}}{I_{6.7}}$ with increasing $\Gamma$. This trend likely arises because larger $\Gamma$ values imply lower plasma temperatures, resulting in a smaller proportion of {Fe\,\textsc{xxvi}} relative to {Fe\,\textsc{xxv}} line emission. 

We also searched for a \gaia counterpart within a 3\arcsec distance from \xmm positions to estimate the distance to the sources using the \gaia parallax. If a counterpart was found, we used the Gaia source ID to determine the distance to the source from \citet{bailer-jones2021}. The distances to the sources with identified \gaia counterparts, along with the corresponding 2--10 keV X-ray luminosities, are provided in Table \ref{table:list_source}.

\section{Discussion}
The characteristics of X-ray point sources towards the GC and Galactic disc, and their relative contributions to the GCXE and GRXE, are still unclear. Recently, \citet{koyama2024} fitted the GRXE spectra with an assembly of X-ray active stars and compact X-ray objects, which revealed a significant excess at the K$\alpha$, He$\alpha$, and Ly$\alpha$ lines. These excesses are interpreted as arising from collisional ionisation equilibrium plasma, which is considered an additional emission component from supernova remnants. Furthermore, \citet{warwick2011} studied the population of faint X-ray sources using \xmm in the Galactic plane, far from the GC ($315^{\circ}<l<45^{\circ}$). They found that faint X-ray sources can be roughly divided into two categories: soft sources with $\Gamma\sim2.5$ and hard sources with $\Gamma\sim1$. The average spectral and Fe-line properties of the hard sources are consistent with the mCVs, and the soft sources are likely associated with relatively nearby coronally active late-type stars, which are identified as bright near-infrared objects within the X-ray positions \citep{warwick2014a}. Subsequently, \citet{warwick2014b} constructed the X-ray luminosity function of the soft and hard sources and found that, in the 6--10 keV band,  80\% of the GRXE intensity originates from point sources, the remainder attributable to X-ray scattering in the interstellar medium and/or young Galactic source populations. Recently, \citet{schmitt2022} attempted to classify sources in the GC by cross-matching \chandra sources with the \gaia EDR3 catalogue, identifying only one third of X-ray point sources, and concluded that the GRXE is produced by optically faint mCVs and DNe. From previous studies \citep{revnivtsev2009,hong2012,warwick2014b,schmitt2022}, it is apparent that in the 6--7 keV band, the vast majority of the GRXE can be resolved into point sources, with mCVs as the primary contributors. The EWs of the 6.4 and 6.9 keV lines of mCVs appear consistent with the GRXE. However, the $\rm EW_{6.7}$ of mCVs in the solar neighbourhood is a factor of 2 to 3 times smaller than that of the GRXE. \citet{xu2016} compared the $\rm EW_{6.7}$ and $\frac{I_{6.9}}{I_{6.7}}$ of the GRXE with different classes of known bright sources in the solar neighbourhood, including mCVs (IPs and polars), non-mCVs such as DNe, and ABs. They concluded that all source classes except DNe show average Fe line diagnostics ($\rm EW_{6.7}$ and $\frac{I_{6.9}}{I_{6.7}}$) significantly different from those observed in the GRXE (see Fig. \ref{fig:ew2_line_ratio}). Thus, without a major contribution from DNe, no combination of other source classes appears to be able to explain the observed $\rm EW_{6.7}$ and $\frac{I_{6.9}}{I_{6.7}}$ of the GRXE. Therefore, they suggested that the GRXE mostly consists of numerous faint DNe, with mCVs only responsible for the high-flux population of the GRXE. From our Fig. \ref{fig:ew2_line_ratio} it is evident that in our sample, numerous sources have $\rm EW_{6.7}$ similar or higher than that of the GCXE or GRXE. This indicates that the discrepancy in $\rm EW_{6.7}$ between various types of CV populations and the GRXE noted by \citet{xu2016} can be explained by incorporating these sources into our sample.

\begin{figure}[]
\centering
\includegraphics[width=\figsize\textwidth]{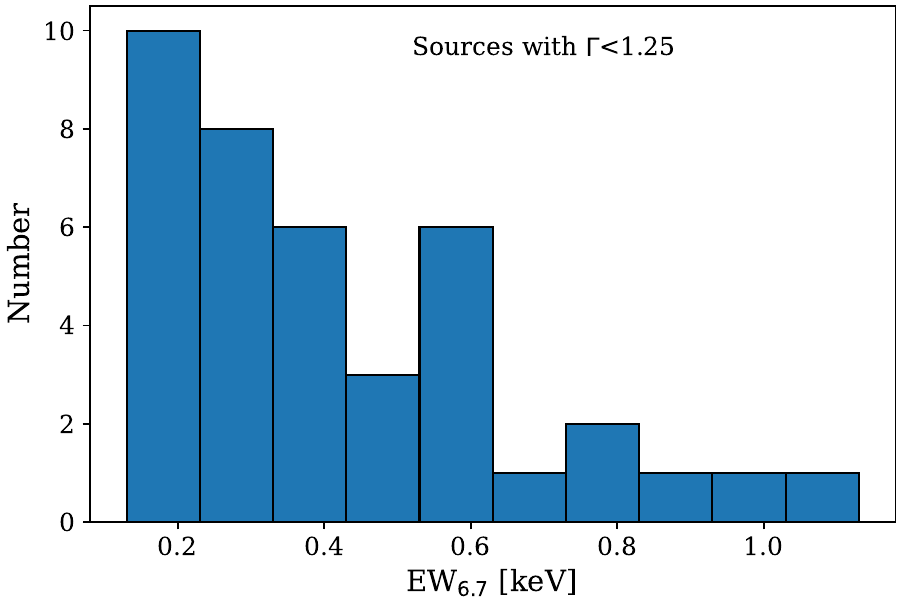}
\caption{Distribution of $\rm EW_{6.7}$ for identified CVs and candidate CVs characterised by $\Gamma<1.25$.}
\label{fig:gamma_ew2_hist}
\end{figure}
\begin{figure}[]
\centering
\includegraphics[width=\figsize\textwidth]{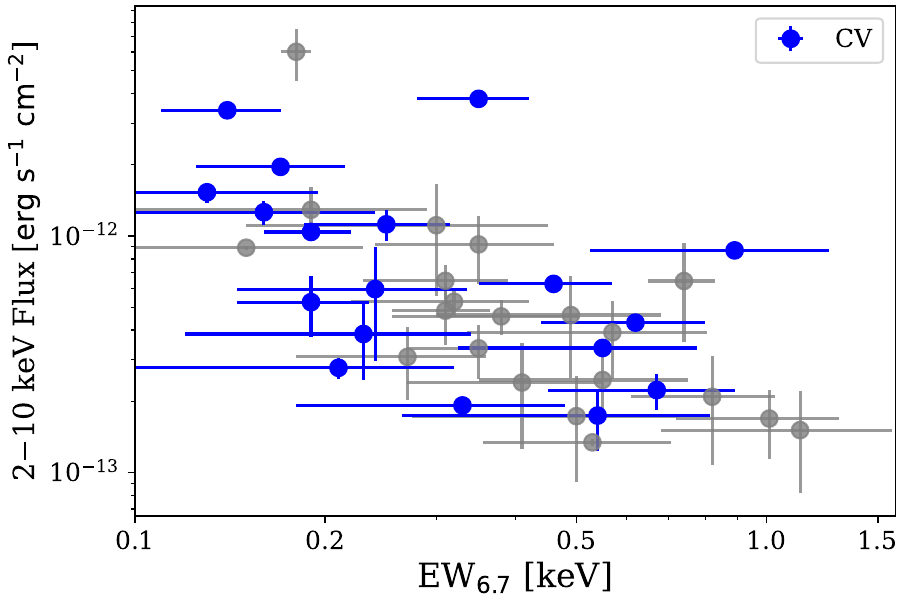}
\caption{2--10 keV flux vs $\rm EW_{6.7}$ for sources with $\Gamma<1.25$. Blue points indicate sources identified as CVs and grey points represent unidentified CV candidates.}
\label{fig:flux_ew2_hard}
\end{figure}
In Fig. \ref{fig:gamma_line_ratio}, we see that most of the identified CVs have smaller values of $\Gamma$, so the unidentified sources with $\Gamma<1.25$ are strong candidates for CVs. There are 21 unidentified sources with $\Gamma<1.25$. They are most likely magnetic-type CVs as their $\frac{I_{6.9}}{I_{6.7}}$ values are similar to those of the mCVs, which are much larger than values found in non-mCVs. Among the 21 CV candidates, most have values of $\frac{I_{6.9}}{I_{6.7}}>0.4$, except for two sources. Non-mCVs, such as quiescent DNe, have weak 6.9 keV line emission (<$\rm EW_{6.9}$>$=95\pm18.6$ eV; \citealt{xu2016}). Therefore, quiescent DNe have a typical <$\frac{I_{6.9}}{I_{6.7}}$>$=0.27\pm0.06$ \citep{xu2016}, much lower than the values found in mCVs such as IPs (<$\frac{I_{6.9}}{I_{6.7}}$>$=0.71\pm0.04$) and polars (<$\frac{I_{6.9}}{I_{6.7}}$>$=0.44\pm0.14$), as illustrated in Fig. \ref{fig:ew2_line_ratio}. Furthermore, in our sample of 18 identified CVs, 13 are classified as mCVs by \citet{mondal2024a} using spin detection, and for only five is it unknown whether they are magnetic or non-magnetic. However, among the five sources, four have distance measurements from \gaia, which indicate luminosities above $10^{33}\rm\ erg\ s^{-1}$, similar to luminosities seen in mCVs; non-mCVs typically have luminosities below $10^{32}\rm\ erg\ s^{-1}$ \citep{suleimanov2022}. The distribution of $\rm EW_{6.7}$ for the CV sample is shown in Fig. \ref{fig:gamma_ew2_hist}. The $\rm EW_{6.7}$ distribution shows a peak around $\sim0.2$ keV and an extended tail up to 1 keV. The CV sample has <$\rm EW_{6.7}$>$=415\pm39$ eV, where the error is the standard error of the sample mean. The peak at $\sim0.2$ keV is likely associated with the typical $\rm EW_{6.7}$ values observed in nearby mCVs. 

\begin{figure}[]
\centering
\includegraphics[width=\figsize\textwidth]{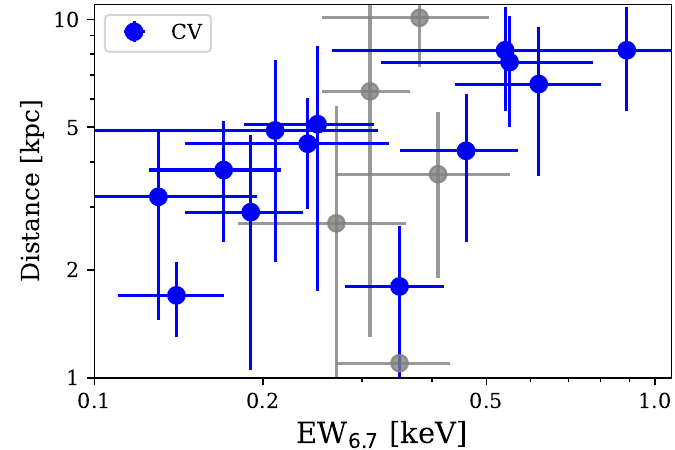}
\includegraphics[width=\figsize\textwidth]{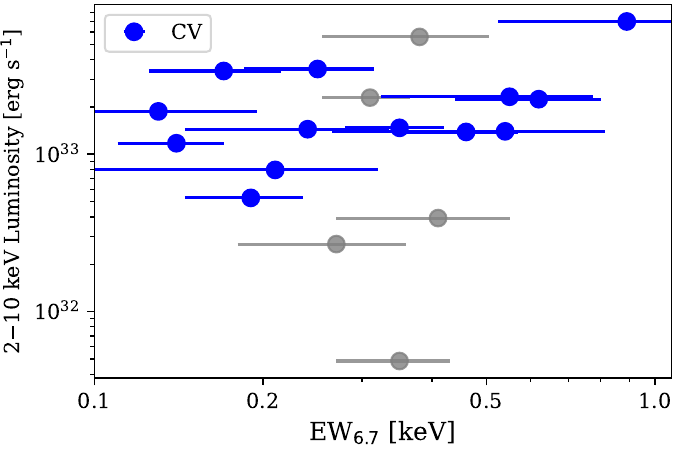}
\caption{Distance (top panel) and 2--10 keV luminosity (bottom panel) vs $\rm EW_{6.7}$ for sources with $\Gamma<1.25$ for which a distance estimation was possible using \gaia parallax. Blue points indicate sources identified as CVs, while grey points represent unidentified CV candidates.}
\label{fig:dist_lumi_ew2_hard}
\end{figure}

The <$\rm EW_{6.7}$> of our CV sample is more than twice as high as the typical $\rm EW_{6.7}$ value found in CVs from the solar neighbourhood. There is a possibility that such a high mean value might be due to selection bias, since mean is computed from a sample of sources with {Fe\,\textsc{xxv}} line detection significance above $2\sigma$. This could result in a bias such that, for faint sources, only the strongest {Fe\,\textsc{xxv}} lines are selected due to low S/N, as the majority of our sources in the CV sample (almost 75\%) have 2--10 keV flux below $10^{-12}\rm \ erg\ s^{-1}\ cm^{-2}$. This may lead to higher <$\rm EW_{6.7}$> because only high $\rm EW_{6.7}$ sources cross the detection threshold above $2\sigma$. In fact, when we plot the 2--10 keV flux versus $\rm EW_{6.7}$ for all our sources with $\Gamma<1.25$, we observe a negative correlation (Fig. \ref{fig:flux_ew2_hard}) suggesting that only the fainter sources with larger $\rm EW_{6.7}$ are selected due to the choice of detection threshold. If there is indeed a selection bias, then one might also expect to detect bright sources with 2--10 keV flux above $10^{-12}\rm \ erg\ s^{-1}\ cm^{-2}$ and $\rm EW_{6.7}>0.4$ keV, which are not present in Fig. \ref{fig:flux_ew2_hard}. This indicates that the inverse trend of flux versus $\rm EW_{6.7}$ is likely not due to selection bias. This becomes clearer when we stack the spectra of all point sources detected in the 6.5--7 keV band and measure the $\rm EW_{6.7}$ of the stacked spectra. The inverse trend of flux versus $\rm EW_{6.7}$ of the hard X-ray sources is likely driven by their distance, as indicated by the positive correlation between distance versus $\rm EW_{6.7}$, shown in the top panel of Fig. \ref{fig:dist_lumi_ew2_hard}. Furthermore, the bottom panel of Fig. \ref{fig:dist_lumi_ew2_hard} shows that $\rm EW_{6.7}$ is independent of luminosity. For solar neighbourhood CVs (including mCVs and non-mCVs) the luminosity versus $\rm EW_{6.7}$ displays an inverse trend, driven by the fact that non-mCVs, such as DNe, have much lower luminosity (typically below $10^{32}\rm\ erg\ s^{-1}$) and much higher $\rm EW_{6.7}$ than mCVs, such as IPs and polars. This is illustrated in Fig. 7 of \citet{xu2016}. The bottom panel of Fig. \ref{fig:dist_lumi_ew2_hard} shows that all sources, except one, have 2--10 keV X-ray luminosities above $2\times10^{32}\rm\ erg\ s^{-1}$,  consistent with the luminosities of mCVs \citep[IPs and polars;][]{suleimanov2022}, rather than non-mCVs, which typically have X-ray luminosities a factor of 10 to 100 times lower than those of mCVs. In addition, these hard sources have a line ratio $\frac{I_{6.9}}{I_{6.7}}>0.4$, much higher than non-mCVs, which suggests that most of these sources are likely mCVs.

\begin{figure*}[]
\centering
\includegraphics[width=\figsizee\textwidth]{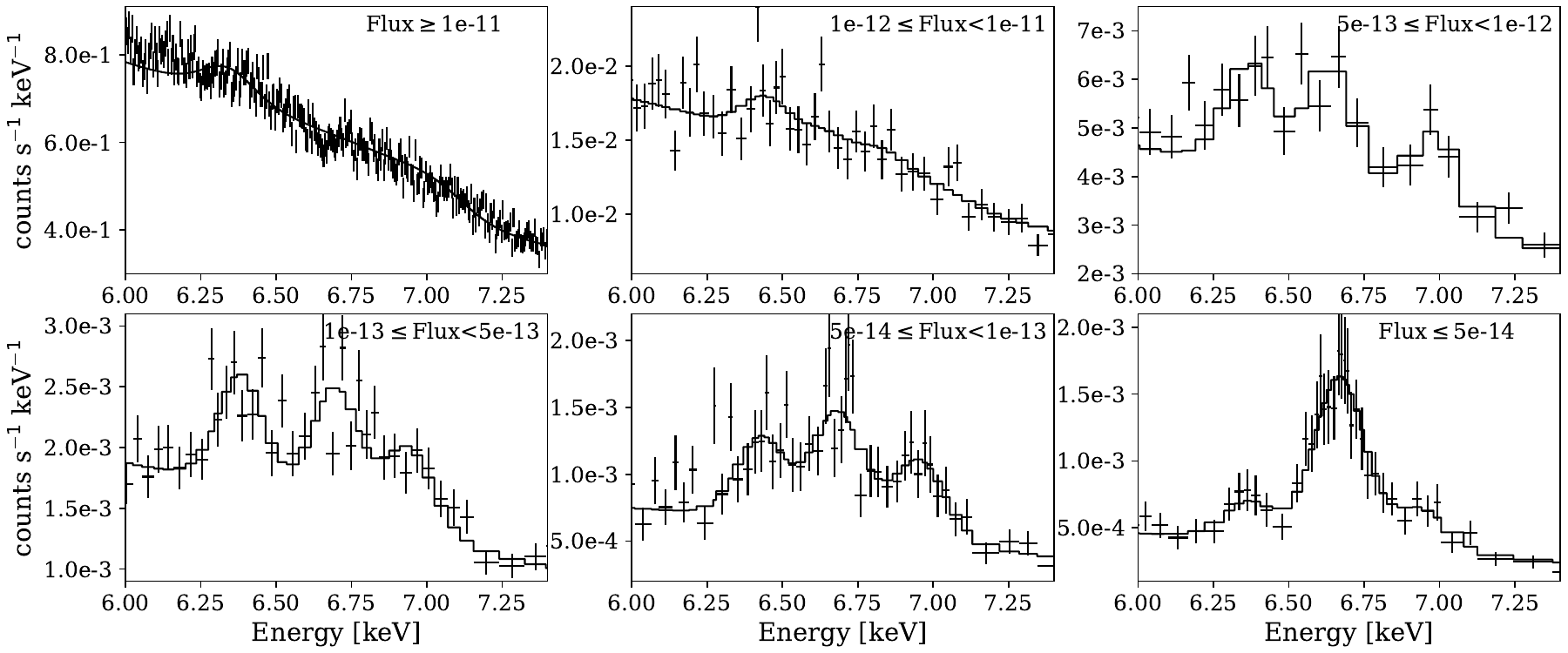}
\caption{Stacked spectra of point sources with HR>-0.2 and ML>14 in the 6.5--7 keV band for different flux groups. The spectra were fitted by a power-law continuum and three Gaussian lines at 6.4, 6.7, and 6.9 keV for the Fe complex.}
\label{fig:spec}
\end{figure*}
We computed the stacked spectra of all sources detected in the 6.5--7 keV band with ML>14. We then stacked the spectra of the sources according to different groups based on their 2--10 keV flux. To separate the two populations of sources, we defined a cut in hardness ratio (HR), with $\rm HR>-0.2$ and $\rm HR\leq-0.2$, for the hard ($\Gamma<1.25$) and soft ($\Gamma\geq1.25$) sources, respectively. The hardness ratio is defined as $\rm HR=\frac{C_{4.5-12\ keV}-C_{1-4.5\ keV}}{C_{4.5-12\ keV}+C_{1-4.5\ keV}}$, where C denotes the net count in the respective energy band. Figure \ref{fig:spec} shows the stacked spectra of sources with $\rm HR>-0.2$ for different flux groups. The stacked spectra are fitted with a power-law continuum and three Gaussians at 6.4, 6.7, and 6.9 keV for the Fe complex. The stacked spectra of bright sources with 2--10 keV flux $>10^{-12}\rm\ erg\ s^{-1}\ cm^{-2}$ do not show the presence of ionised {Fe\,\textsc{xxv}} and {Fe\,\textsc{xxvi}} lines; however, there is a weak indication that the neutral 6.4 keV line is present. The Fe complex becomes more prominent in the stacked spectra of sources with 2--10 keV flux $<10^{-12}\rm\ erg\ s^{-1}\ cm^{-2}$. From Fig. \ref{fig:spec} it is clear that the strength of the {Fe\,\textsc{xxv} line relative to the neutral Fe $K_{\alpha}$ line and the {Fe\,\textsc{xxvi} line increases as we stack sources with fainter flux. Figure \ref{fig:flux_ew2} shows the $\rm EW_{6.7}$ obtained from the stacked spectra as a function of the 2--10 keV mean flux, suggesting that the faint sources have significantly larger $\rm EW_{6.7}$ compared to the brighter sources, as seen in Fig. \ref{fig:flux_ew2_hard}. This ensures that the large <$\rm EW_{6.7}$> of our identified and candidate CV sources is less likely due to a selection bias, but is due to the fact that faint sources are more likely to have higher $\rm EW_{6.7}$.
\begin{figure}[]
\centering
\includegraphics[width=\figsize\textwidth]{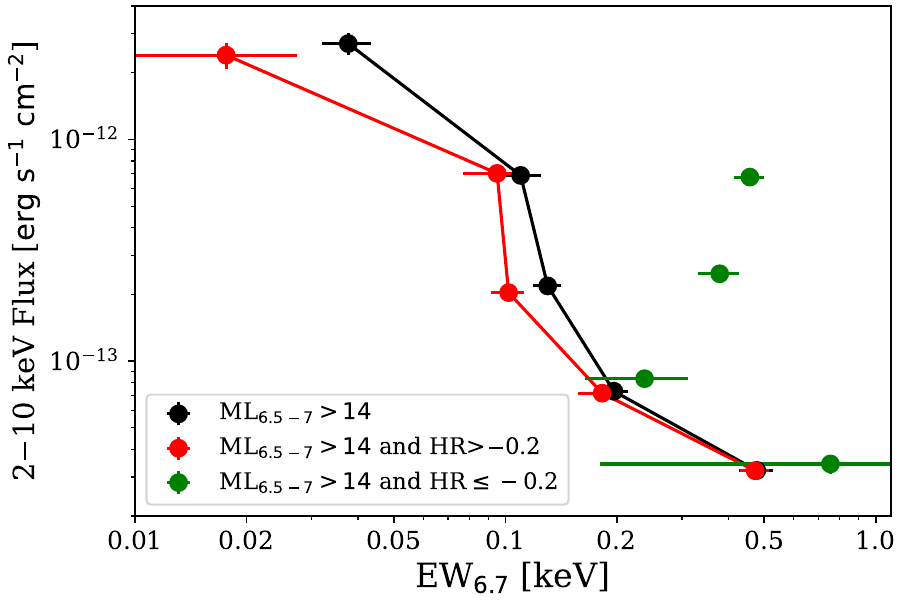}
\caption{Stacked-spectra $\rm EW_{6.7}$ for point sources in grouped by 2--10 keV flux. All sources detected with ML>14 in the 6.5--7 keV band are considered (black line-connected dots). Red and green points correspond to sources with $\rm HR>-0.2$ and $\rm HR\leq-0.2$, respectively. The 2--10 keV observed flux represents the mean flux of each group, and the error bars show the standard error of the mean. The $\rm EW_{6.7}$ errors are at the $1\sigma$ confidence level obtained from {\sc xspec}.}
\label{fig:flux_ew2}
\end{figure}

The extended tail of our CV candidate sample, leading to a higher <$\rm EW_{6.7}$>$=415\pm39$ eV, can be explained by assuming that they have a metallicity higher than solar. For example, in the solar neighbourhood, a typical CV with a plasma temperature of 15 keV has an $\rm EW_{6.7}$ of $\sim0.14$ keV. However, for the same plasma temperature, assuming an iron abundance of 2.5 times solar yields an $\rm EW_{6.7}$ of 0.3 keV. Conversely, for a plasma temperature of 10 keV, $\rm EW_{6.7}=0.35$ and 0.72 keV for abundance values 1.0, and 2.5, respectively, as shown in Fig. \ref{fig:spec_sim} through spectral simulation. Figure \ref{fig:spec_sim} indicates that the large <$\rm EW_{6.7}$> of 415 eV can be explained by assuming a relatively low plasma temperature of 10--12 keV and an abundance of 1--2 times solar. At higher plasma temperatures, this will require a significantly higher abundance, which is unlikely.
\begin{figure}[]
\centering
\includegraphics[width=\figsize\textwidth]{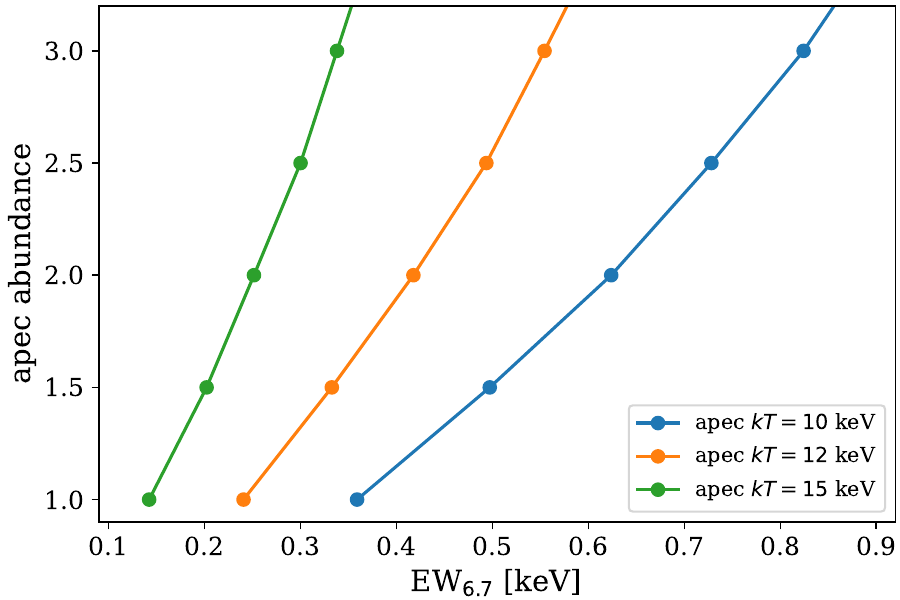}
\caption{Dependence of $\rm EW_{6.7}$ for three different plasma temperatures: 10, 12, and 15 keV. Dots indicate abundance values used in spectral simulations with an \texttt{apec} model, and
the $\rm EW_{6.7}$ values were estimated by fitting the simulated spectra using a model composed of power-law plus Gaussian at 6.7 keV
with width fixed to zero eV.}
\label{fig:spec_sim}
\end{figure}

We detected a significant number of sources that emit a strong 6.7 keV line and have a very soft X-ray spectrum. Figure \ref{fig:apec_abun_ew2_hist} shows the $\rm EW_{6.7}$ distribution for these soft sources. The mean $\rm EW_{6.7}$, with standard error,  is <$\rm EW_{6.7}$>$=1.0\pm0.1$ keV. We identified counterparts for 13 soft sources within a distance of 3\arcsec\ from the \xmm position. Two are associated with RS CVn-type stars, a class of variable binary stars known to emit strong coronal X-ray emission and 6.7 keV line emission \citep{audard2001}. The X-ray continuum emission from these types of sources is typically described by two temperature components: a line-emitting plasma with a characteristic temperature of $\sim0.5$ keV, and a Bremsstrahlung continuum of 2--8 keV \citep{swank1981}. The RS CVn ABs show transient behaviour in the {Fe\,\textsc{xxv}} line; however, the $\rm EW_{6.7}$ is much larger during flare events \citep{mewe1997,gudel1999,inoue2024}. Furthermore, five of the 13 sources with counterparts are classified as young stellar objects (YSOs), and six are massive coronally active stars. It is therefore difficult to determine the true nature of the sources with soft X-ray spectra ($\Gamma\geq1.25$) and large $\rm EW_{6.7}$, as this combination of spectral properties is observed across several different source classes. Active binary stars such as RS CVn and coronally active stars such as YSOs show a strong 6.7 keV line; however, their X-ray luminosities are much lower ($<10^{32}\rm\ erg\ s^{-1}$). Massive WR-OB binary stars also emit $\rm EW_{6.7}>1$ keV, originating from wind-wind collisions, with 2--10 keV luminosities reaching up to $10^{35}\rm\ erg\ s^{-1}$. It is also not uncommon for polars to have a similar spectrum shape of $\Gamma>1.25$, while IPs typically exhibit much harder spectra. For example, \xmm detected such a system toward the GC (XMM J174544--2913.0) with $\Gamma=2.00^{+0.13}_{-0.27}$ and $\rm EW_{6.7}=2.4^{+0.4}_{-0.5}$ keV \citep{sakano2005}. Many polar-type systems detected by \emph{ASCA} also exhibit $\Gamma\sim2$ and $\rm EW_{6.7}$ above 1 keV (AX J2315--592, \citealt{misaki1996}; RX J1802.1+1804, \citealt{ishida1998}; AX J1842.8--0423, \citealt{terada1999}). The nature of XMM J174544--2913.0 and AX J1842.8--0423 remains unclear due to their transient X-ray emission; however, the two other systems are confirmed polars. Typically, mCVs such as polars in the solar neighbourhood have much lower $\rm EW_{6.7}$. \citet{terada2004} proposed that the large $\rm EW_{6.7}$ observed in polars detected by \emph{ASCA},  results from collimation of {Fe\,\textsc{xxv}} line photons along the accretion column axis due to resonance scattering, which may occur in the case of a pole-on inclination.
\begin{figure}[]
\centering
\includegraphics[width=\figsize\textwidth]{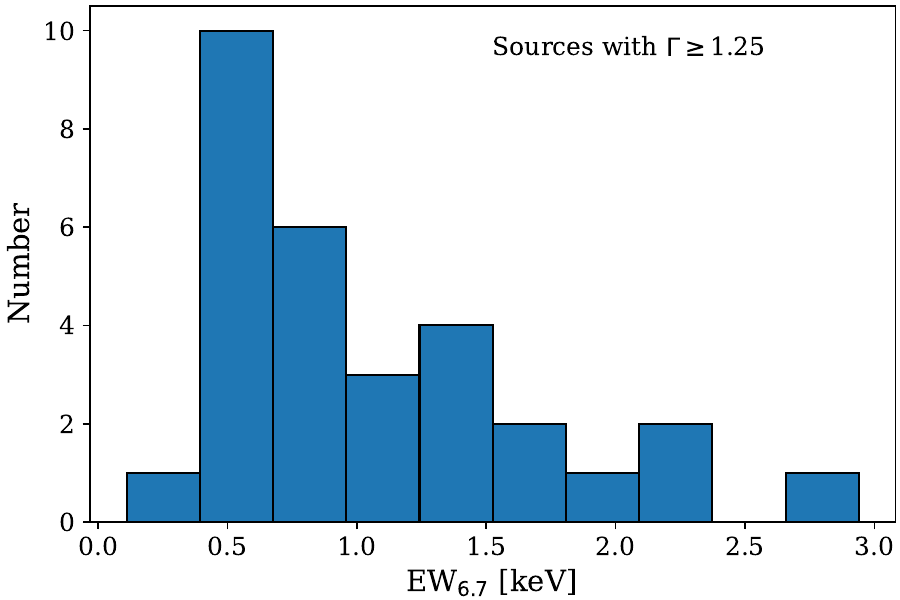}
\caption{Distribution of $\rm EW_{6.7}$ for identified stellar-type sources and unidentified sources with $\Gamma\geq1.25$.}
\label{fig:apec_abun_ew2_hist}
\end{figure}

We compared the <$\rm EW_{6.7}$> and <$\frac{I_{6.7}}{I_{6.9}}$> of the hard ($\Gamma<1.25$) and soft sources ($\Gamma\geq1.25$) with the GCXE and GRXE. The luminosity inferred using \gaia distances and the line ratio $\frac{I_{6.9}}{I_{6.7}}$ of the hard sources suggest that they are likely to be mCVs; however, the <$\rm EW_{6.7}$> of these sources is significantly higher than that of solar neighbourhood mCVs and approaches values seen in the GCXE and GRXE. This indicates that the GCXE and GRXE can have significant contributions from candidate mCVs with large $\rm EW_{6.7}$. This finding contrasts with earlier studies that proposed non-mCVs as the major contributors to the GRXE, based on comparisons of the Fe emission line properties of sources in the solar neighbourhood with those observed in the GRXE \citep{xu2016,schmitt2022}. The sensitivity limit for point source detection in our survey is around $\sim10^{-14}\ \rm erg\ s^{-1}\ cm^{-2}$. Furthermore, almost $\sim$50\% of our sample in the 6.5--7 keV band consists of hard sources, most of which are likely mCVs. This indicates that among the sources with a flux of $10^{-14}\ \rm erg\ s^{-1}\ cm^{-2}$ or greater that contribute to the GDXE in the 6.5--7 keV band, nearly half originate from mCVs,  with the remainder contributed by soft sources such as non-mCVs, ABs, YSOs, and coronally active stars. In addition, IPs are likely to contribute to the higher flux level of $\sim8\times10^{-14}\ \rm erg\ s^{-1}\ cm^{-2}$ or higher. Figure \ref{fig:mean_ew2_line_ratio} shows the sample mean and $1\sigma$ standard deviation. The <$\rm EW_{6.7}$> and <$\frac{I_{6.7}}{I_{6.9}}$> measurements for both the CV population and soft sources are consistent with those of the GCXE within $1\sigma$. This indicates that the GCXE can be explained by the hard CV population,  the soft sources, or a combination of both. On the other hand, in contrast to the GCXE, the GRXE exhibits a much lower $\frac{I_{6.7}}{I_{6.9}}$ than the CV population. However, the line intensity ratio $\frac{I_{6.7}}{I_{6.9}}$ of the GRXE is consistent within the $1\sigma$ measurement of the soft sources. This indicates that low flux ($<1.8\times10^{-13}\rm\ erg\ s^{-1}\ cm^{-2}$) soft sources, such as ABs and coronally active stars, may contribute a significant amount to the GRXE.
\begin{figure}[]
\centering
\includegraphics[width=\figsize\textwidth]{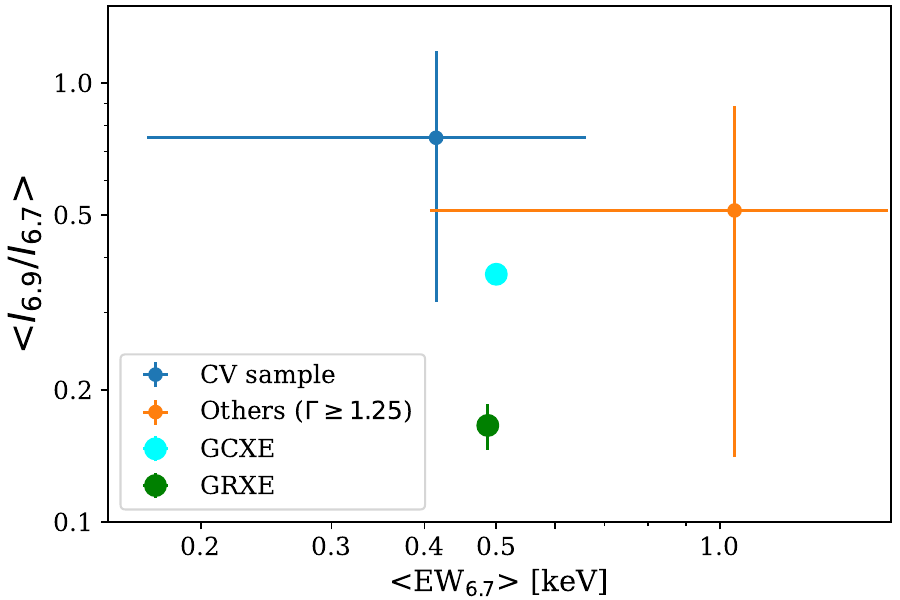}
\caption{Comparison of the mean $\rm EW_{6.7}$ and $\frac{I_{6.9}}{I_{6.7}}$ for our hard CV and soft source samples with those of the GCXE and GRXE. The measurement uncertainties represent the $1\sigma$ standard deviation. The GCXE and GRXE values are taken from \citet{nobukawa2016}.}
\label{fig:mean_ew2_line_ratio}
\end{figure}

\section{Conclusions}
We aimed to characterise the sources contributing to the Galactic diffuse X-ray emission in the 6.7 keV band. The sample of sources contributing to this Galactic diffuse X-ray emission consists primarily of CVs, coronally active stars, and ABs. Previous studies suggested that the 6.7 keV line emission of the Galactic diffuse X-ray emission originates primarily from mCVs. However, the $\rm EW_{6.7}$ and line intensity ratio $\frac{I_{6.9}}{I_{6.7}}$ of the GCXE and GRXE do not match those of the mCVs. To address this discrepancy, we measured the X-ray spectral slope $\Gamma$, $\rm EW_{6.7}$, and $\frac{I_{6.9}}{I_{6.7}}$ for a substantial number of individual sources in the inner Galactic disc. The main conclusions of this study are as follows.
\begin{itemize}
    \item We detected the {Fe\,\textsc{xxv}} line above the $2\sigma$ confidence level in a sample of 72 sources. The $\rm EW_{6.7}$ ranges from a few hundred eV to $\sim3$ keV. The sample consists primarily of two populations: hard and soft sources, characterised by an X-ray spectral slope $\Gamma$ below and above 1.25, respectively.
    \item The $\rm EW_{6.7}$ correlates with the X-ray spectral slope, $\Gamma$. The harder sources with smaller values of $\Gamma$ have a lower $\rm EW_{6.7}$ line than the softer sources with large $\Gamma$.
    \item The vast majority of hard sources are identified with known CVs. The X-ray spectral slope $\Gamma$ and the line intensity ratio $\frac{I_{6.9}}{I_{6.7}}$ of the subsample suggest that they are prime candidates for CV. The <$\rm EW_{6.7}$> of these harder CV-type sources is $415\pm39$ eV, significantly higher than the typical $\rm EW_{6.7}$ value of $\sim170$ eV found in CVs in the solar neighbourhood. This suggests that the iron abundance of these harder sources must be 1--2 times the solar value, with a plasma temperature of 10--12 keV.
    \item The soft source sample is associated with stellar-like activities and has an <$\rm EW_{6.7}$>$=1.1\pm0.1$ keV. These sources have a mean spectral index $\Gamma\sim1.8$, and their <$\rm EW_{6.7}$> values are consistent with those observed in coronally active stars, AB-like RS CVn stars, and YSOs.
    \item Galactic centre X-ray emission can be explained by a mixture of CVs with higher solar abundance and a contribution from soft sources. However, the soft sources may contribute significantly to Galactic ridge X-ray emission.
\end{itemize}


\begin{acknowledgements}
SM, GP, and TB acknowledge support from Bando per il Finanziamento della Ricerca Fondamentale 2022 dell'Istituto Nazionale di Astrofisica (INAF): GO Large program and from the Framework per l'Attrazione e il Rafforzamento delle Eccellenze (FARE) per la ricerca in Italia (R20L5S39T9). GP also acknowledges financial support from the European Research Council (ERC) under the European Union's Horizon 2020 research and innovation program HotMilk (grant agreement No. 865637). MM gratefully acknowledges support for this work from NASA ADAP grant 80NSSC24K0639. We thank the referee
for the constructive comments, corrections, and suggestions that significantly improved the manuscript.

\end{acknowledgements}

%
%

\bibliographystyle{aa} 
\bibliography{refs}

\begin{appendix}


\begin{table*}
\section{Additional tables}
\caption{The various details of the X-ray sources emitting {Fe\,\textsc{xxv}} line.}
\label{table:list_source}
\setlength{\tabcolsep}{2.5pt}                   
\renewcommand{\arraystretch}{1.5}               
\centering
\begin{tabular}{c c c c c c c c c c c c c c}
\hline\hline
XMM ID & RA & DEC & $N_{\rm H}$ & $\Gamma$ & $\rm EW_{6.7}$ & $I_{6.7}$ & $I_{6.9}$ & $F_{\rm 2-10\ keV}$ & Type & Dist & $L_{\rm X}$\\ \hline

0402040101 & 260.2802 & -37.4527 & $3.55^{+0.13}_{-0.12}$ & $5.27^{+0.14}_{-0.14}$ & $2.94^{+0.41}_{-0.32}$ & $10.9^{+1.5}_{-1.5}$ & $6.69^{+1.77}_{-1.77}$ & $20.3^{+0.6}_{-0.6}$ \\ \hline
0886070501 & 260.5310 & -35.5907 & $5.22^{+0.69}_{-0.62}$ & $1.69^{+0.23}_{-0.21}$ & $0.73^{+0.14}_{-0.17}$ & $7.64^{+1.53}_{-1.53}$ & $6.44^{+1.45}_{-1.45}$ & $8.85^{+0.05}_{-1.28}$ & YSO\\ \hline
0405640701 & 261.4961 & -36.2176 & $18.05^{+14.70}_{-10.21}$ & $1.38^{+1.58}_{-1.17}$ & $0.55^{+0.26}_{-0.22}$ & $7.40^{+3.76}_{-3.76}$ & $6.18^{a}$ & $8.01^{+1.12}_{-0.20}$  \\ \hline
0150220101 & 262.0124 & -35.0110 & $0.25^{+0.02}_{-0.02}$ & $1.87^{+0.06}_{-0.06}$ & $0.59^{+0.13}_{-0.11}$ & $2.54^{+0.74}_{-0.74}$ & $1.76^{+0.84}_{-0.84}$ & $4.82^{+0.15}_{-0.16}$ & YSO & $0.45^{+0.03}_{-0.03}$ & $1.17\times10^{31}$\\ \hline
0822210101 & 262.6283 & -34.5389 & $0.221^{+0.004}_{-0.003}$ & $2.05^{+0.02}_{-0.02}$ & $0.83^{+0.06}_{-0.07}$ & $49.4^{+3.8}_{-3.9}$ & $14.5^{+2.6}_{-2.6}$ & $70.6^{+0.4}_{-0.2}$ & RS CVn & $0.38^{+0.01}_{-0.01}$ & $1.22\times10^{32}$\\ \hline
0916801401 & 262.6390 & -33.6543 & $0.39^{+0.01}_{-0.01}$ & $3.30^{+0.06}_{-0.06}$ & $2.11^{+0.47}_{-0.38}$ & $5.74^{+1.14}_{-1.13}$ & $2.45^{+1.08}_{-1.08}$ & $6.29^{+0.47}_{-0.42}$ & RS CVn & $0.35^{+0.01}_{-0.01}$& $9.22\times10^{30}$\\ \hline
0743610101 & 262.7402 & -33.5136 & $0.38^{+0.01}_{-0.01}$ & $2.18^{+0.04}_{-0.04}$ & $0.40^{+0.10}_{-0.10}$ & $1.62^{+0.41}_{-0.41}$ & $0.52^{+0.29}_{-0.29}$ & $4.96^{+0.03}_{-0.02}$ && $0.76^{+0.2}_{-0.1}$ & $3.43\times10^{31}$\\ \hline
0861171201 & 262.7453 & -35.1365 & $1.07^{+0.48}_{-0.21}$ & $0.43^{+0.22}_{-0.21}$ & $0.13^{+0.06}_{-0.07}$ & $2.51^{+1.34}_{-1.34}$ & $1.47^{a}$ & $15.3^{+1.1}_{-1.9}$ & CV$^{b}$ & $3.2^{+2.2}_{-1.3}$ & $1.87\times10^{33}$\\ \hline
0701230701 & 263.8276 & -32.9082 & $2.23^{+0.26}_{-0.24}$ & $1.43^{+0.13}_{-0.12}$ & $0.54^{+0.08}_{-0.10}$ & $4.52^{+0.91}_{-0.92}$ & $2.09^{+0.84}_{-0.84}$ & $7.33^{+0.15}_{-0.17}$ & HMXB & $5.4^{+3.0}_{-2.9}$ & $2.56\times10^{33}$\\ \hline
0886030601 & 263.9425 & -32.1192 & $0.30^{+0.04}_{-0.04}$ & $2.12^{+0.14}_{-0.13}$ & $2.08^{+0.96}_{-0.64}$ & $2.49^{+0.96}_{-0.96}$ & $1.72^{+0.93}_{-0.94}$ & $1.60^{+0.17}_{-0.15}$ && $1.1^{+0.03}_{-0.04}$ & $2.32\times10^{31}$\\ \hline
0886031601 & 264.4924 & -32.9128 & $0.93^{+2.00}_{-0.15}$ & $-0.73^{+0.59}_{-0.43}$ & $1.01^{+0.25}_{-0.33}$ & $1.98^{+0.52}_{-0.52}$ & $1.04^{+0.69}_{-0.68}$ & $1.69^{+0.33}_{-0.78}$\\ \hline
0886020101 & 264.6545 & -30.8050 & $0.67^{+0.33}_{-0.25}$ & $0.72^{+0.22}_{-0.21}$ & $0.33^{+0.17}_{-0.13}$ & $0.79^{+0.45}_{-0.45}$ & $0.60^{a}$ & $1.92^{+0.11}_{-0.03}$ & CV$^{b}$\\ \hline
0764191601 & 264.7088 & -28.8026 & $3.74^{+3.65}_{-1.87}$ & $0.22^{+0.48}_{-0.38}$ & $0.62^{+0.17}_{-0.19}$ & $3.50^{+0.95}_{-0.95}$ & $2.69^{+0.91}_{+0.90}$ & $4.30^{+0.16}_{-0.22}$ & CV & $6.6^{+3.4}_{-2.5}$ & $2.24\times10^{33}$\\ \hline
0886121201 & 264.8833 & -30.4118 & $3.09^{+1.13}_{-0.93}$ & $0.19^{+0.26}_{-0.25}$ & $0.15^{+0.08}_{-0.08}$ & $1.84^{+0.98}_{-0.98}$ & $1.89^{+1.15}_{-1.15}$ & $8.92^{+0.17}_{-0.21}$ \\ \hline
0554720101 & 264.8980 & -30.3452 & $1.55^{+0.61}_{-0.46}$ & $1.94^{+0.46}_{-0.39}$ & $1.35^{+0.49}_{-0.47}$ & $1.81^{+0.51}_{-0.51}$ & $1.20^{+0.37}_{-0.37}$ & $1.9^{+0.46}_{-0.55}$ && $2.6^{+2.2}_{-0.7}$ & $1.54\times10^{32}$\\ \hline
0764191101 & 265.0377 & -28.7907 & $0.70^{+0.14}_{-0.13}$ & $0^{+0.09}_{-0.08}$ & $0.35^{+0.07}_{-0.07}$ & $17.0^{+3.5}_{-3.5}$ & $10.9^{+3.5}_{-3.5}$ & $38.0^{+0.6}_{+0.5}$ & CV$^{b}$ & $1.8^{+1.2}_{-0.5}$ & $1.47\times10^{33}$\\ \hline
0764191201 & 265.0666 & -29.0605 & $0.45^{+0.04}_{-0.04}$ & $0.62^{+0.04}_{-0.04}$ & $0.14^{+0.03}_{-0.03}$ & $5.92^{+1.1}_{-1.1}$ & $6.57^{+1.09}_{-1.09}$ & $34.0^{+0.1}_{-0.3}$ & CV$^{b}$ & $1.7^{+0.4}_{-0.4}$ & $1.18\times10^{33}$\\ \hline
0886010601 & 265.1409 & -30.2503 & $0.81^{+0.10}_{-0.09}$ & $0.38^{+0.07}_{-0.06}$ & $0.17^{+0.05}_{-0.04}$ & $4.22^{+1.27}_{-1.27}$ & $3.51^{+1.29}_{-1.29}$ & $19.6^{+0.2}_{-0.2}$ & CV$^{b}$ & $3.8^{+1.5}_{-1.3}$ & $3.39\times10^{33}$\\ \hline
0764191301 & 265.1513 & -28.1447 & $1.54^{+0.84}_{-0.61}$ & $1.86^{+0.64}_{-0.56}$ & $2.11^{+0.54}_{-1.06}$ & $1.72^{+0.61}_{-0.61}$ & $2.72^{+0.98}_{-0.98}$ & $0.98^{+0.11}_{-0.13}$ \\ \hline
0886020201 & 265.2754 & -31.1925 & $3.26^{+0.40}_{-0.36}$ & $2.76^{+0.25}_{-0.23}$ & $0.75^{+0.29}_{-0.30}$ & $2.17^{+0.88}_{-0.88}$ & $0.52^{a}$ & $3.67^{+0.14}_{-0.16}$ & Star & $3.7^{+1.3}_{-1.3}$ & $6.01\times10^{32}$\\ \hline
0801680101 & 265.4268 & -27.9752 & $0.12^{+0.11}_{-0.09}$ & $0.62^{+0.18}_{-0.17}$ & $0.55^{+0.26}_{-0.22}$ & $2.24^{+0.98}_{-0.98}$ & $1.15^{+0.84}_{-0.84}$ & $3.36^{+0.14}_{-0.13}$ & CV & $7.6^{+2.7}_{-2.5}$ & $2.32\times10^{33}$\\ \hline
0764190601 & 265.5703 & -28.6192 & $1.61^{+0.31}_{-0.27}$ & $1.75^{+0.26}_{-0.24}$ & $0.92^{+0.37}_{-0.44}$ & $5.98^{+2.75}_{-2.75}$ & $6.57^{+3.04}_{-3.04}$ & $6.72^{+0.57}_{-1.43}$ && $7.3^{+3.2}_{-2.2}$ & $4.29\times10^{33}$\\ \hline
0862470101 & 265.9788 & -29.5715 & $1.57^{+0.45}_{-0.39}$ & $1.41^{+0.30}_{-0.28}$ & $1.15^{+0.31}_{-0.34}$ & $2.17^{+0.54}_{-0.54}$ & $1.18^{+0.43}_{-0.43}$ & $1.76^{+0.09}_{-0.09}$ & Star & $2.0^{+1.7}_{-1.2}$ & $8.43\times10^{31}$\\ \hline
0821120101 & 266.0488 & -28.8230 & $0.99^{+0.73}_{-0.43}$ & $1.24^{+0.50}_{-0.41}$ & $1.13^{+0.48}_{-0.42}$ & $1.59^{+0.56}_{-0.55}$ & $1.49^{+0.43}_{-0.43}$ & $1.51^{+0.29}_{-1.08}$\\ \hline
0201200101 & 266.1901 & -27.2293 & $0.36^{+0.02}_{-0.01}$ & $1.70^{+0.04}_{-0.04}$ & $0.47^{+0.07}_{-0.12}$ & $5.80^{+1.03}_{-1.03}$ & $1.63^{+0.78}_{-0.78}$ & $12.9^{+5.0}_{-6.0}$ & HMXB$^{b}$ & $1.2^{+0.03}_{-0.03}$ & $2.22\times10^{32}$\\ \hline
0723410501 & 266.2272 & -28.8207 & $4.50^{+1.41}_{-1.18}$ & $-0.17^{+0.28}_{-0.26}$ & $0.18^{+0.09}_{-0.11}$ & $14.0^{+4.5}_{-4.5}$ & $5.85^{+5.83}_{-5.83}$ & $60.1^{+4.2}_{-25.4}$ \\ \hline
0762250301 & 266.2480 & -28.7696 & $4.53^{+5.36}_{-3.20}$ & $1.52^{+1.50}_{-1.16}$ & $0.52^{+0.37}_{-0.42}$ & $0.76^{+0.49}_{-0.49}$ & $0.46^{+0.38}_{-0.38}$ & $1.23^{+0.09}_{-0.15}$ && $1.7^{+0.99}_{-0.44}$ & $4.25\times10^{31}$\\ \hline
0694640201 & 266.2494 & -29.3284 & $10.24^{+9.11}_{-6.41}$ & $0.23^{+0.71}_{-0.67}$ & $0.53^{+0.14}_{-0.21}$ & $1.20^{+0.51}_{-0.50}$ & $0.90^{+0.05}_{-0.05}$ & $1.34^{+0.06}_{-0.05}$ \\ \hline
0862470301 & 266.2503 & -29.2676 & $5.55^{+8.08}_{-4.11}$ & $0.35^{+0.89}_{-0.67}$ & $0.50^{+0.15}_{-0.30}$ & $1.17^{+0.52}_{-0.53}$ & $0.52^{a}$ & $1.73^{+0.10}_{-1.53}$ \\ \hline
0691760101 & 266.2602 & -27.1710 & $2.26^{+1.12}_{-0.87}$ & $0.61^{+0.34}_{-0.32}$ & $0.44^{+0.13}_{-0.15}$ & $1.94^{+0.70}_{-0.71}$ & $0.73^{a}$ & $3.54^{+0.35}_{-0.12}$ & Star & $6.8^{+3.2}_{-2.5}$ & $1.96\times10^{33}$\\ \hline
0865510101 & 266.3207 & -32.2321 & $2.58^{+0.21}_{-0.19}$ & $0.96^{+0.07}_{-0.07}$ & $0.19^{+0.03}_{-0.03}$ & $2.48^{+0.44}_{-0.44}$ & $2.23^{+0.48}_{-0.48}$ & $10.4^{+2.0}_{-5.0}$ & CV$^{b}$\\ \hline
0886090801 & 268.3680 & -24.7742 & $0.46^{+0.2}_{-0.2}$ & $1.36^{+0.03}_{-0.03}$ & $0.11^{+0.04}_{-0.03}$ & $7.38^{+1.6}_{-1.6}$ & $9.75^{+1.76}_{-1.76}$ & $59.5^{+1.2}_{-1.1}$ & Star & $1.8^{+0.3}_{-0.2}$ & $2.31\times10^{33}$\\ \hline	
0764190101 & 266.3694 & -28.1560 & $0.42^{+0.04}_{-0.04}$ & $1.37^{+0.07}_{-0.07}$ & $0.46^{+0.14}_{-0.08}$ & $3.04^{+0.80}_{-0.80}$ & $2.48^{+0.89}_{-0.89}$ & $6.27^{+0.53}_{-0.49}$ & CV & $4.3^{+2.4}_{-1.4}$ & $1.39\times10^{33}$\\ \hline
0202670701 & 266.4004 & -28.9440 & $7.32^{+1.58}_{-1.29}$ & $2.69^{+0.44}_{-0.39}$ & $1.11^{+0.25}_{-0.21}$ & $1.91^{+0.37}_{-0.37}$ & $1.61^{+0.62}_{-0.62}$ & $1.70^{+0.04}_{-1.24}$ & Star\\ \hline
\end{tabular}
\tablefoot{$N_{\rm H}$, $\rm EW_{6.7}$, and $F_{\rm 2-10\ keV}$ are given in units of $10^{22}\rm\ cm^{-2}$, keV, and $10^{-13}$ erg s$^{-1}$ cm$^{-2}$, respectively. $I_{6.7}$ and $I_{6.9}$ are the intensity of {Fe\,\textsc{xxv}} and {Fe\,\textsc{xxvi}} lines given in units of $10^{-6}$ photons cm$^2$ s$^{-1}$ at the line energy. The source type is taken from SIMBAD if a counterpart is found within 3\arcsec\ distance of \xmm position. The $I_{6.9}$ column with $^a$ indicates the values with upper limit measurement and -- is for sources with no data points above 6.9 keV with signal-to-noise ratio above 3. The Type column with $^b$ indicates sources in which the classification was also done by the detection of an X-ray spin period in \citet{mondal2022,mondal2023,mondal2024a,mondal2024b} which suggests the magnetic nature of the sources. The Dist column indicates the distances to the sources obtained from \gaia parallax in units of kpc. The X-ray luminosity in the 2--10 keV band $L_{\rm X}$ is given in units of erg s$^{-1}$.}
\end{table*}

\begin{table*}
\ContinuedFloat
\caption{Continued.}
\setlength{\tabcolsep}{2.5pt}                   
\renewcommand{\arraystretch}{1.5}               
\centering
\begin{tabular}{c c c c c c c c c c c c c c c}
\hline\hline
XMM ID & RA & DEC & $N_{\rm H}$ & $\Gamma$ & $\rm EW_{6.7}$ & $I_{6.7}$ & $I_{6.9}$ & $F_{\rm 2-10\ keV}$ & Type & Dist & $L_{\rm X}$\\ \hline

0694641101 & 266.4156 & -29.0064 & $9.14^{+0.65}_{-0.61}$ & $3.04^{+0.21}_{-0.20}$ & $0.60^{+0.08}_{-0.13}$ & $10.8^{+2.0}_{-2.0}$ & $3.09^{+1.79}_{-1.79}$ & $17.0^{+9.0}_{-5.1}$ \\ \hline
\hline
0202670701 & 266.4332 & -28.9967 & $9.00^{+1.32}_{-1.17}$ & $2.70^{+0.35}_{-0.32}$ & $1.28^{+0.18}_{-0.23}$ & $6.00^{+1.07}_{-1.07}$ & $1.79^{+0.84}_{-0.84}$ & $4.25^{+0.20}_{-1.95}$ \\ \hline
0112971001 & 266.4352 & -29.2170 & $15.27^{+9.08}_{-4.25}$ & $1.83^{+1.54}_{-0.82}$ & $1.76^{+0.31}_{-0.25}$ & $95.3^{+16.8}_{-16.8}$ & $15.7^{+9.7}_{-9.7}$ & $43.5^{+2.3}_{-35.7}$ & \\ \hline
0111350101 & 266.4598 & -28.8202 & $8.52^{+1.03}_{-0.93}$ & $3.25^{+0.34}_{-0.32}$ & $1.40^{+0.24}_{-0.25}$ & $7.64^{+1.38}_{-1.38}$ & $0.88^{+0.65}_{-0.65}$ & $5.90^{+0.31}_{-2.28}$ & Star & $6.2^{+3.6}_{-3.0}$ & $2.71\times10^{33}$\\ \hline
0723410401 & 266.4616 & -28.9083 & $7.92^{+6.63}_{-4.40}$ & $0.91^{+0.91}_{-0.75}$ & $0.55^{+0.19}_{-0.21}$ & $1.83^{+0.63}_{-0.63}$ & $1.56^{+0.63}_{-0.63}$ & $2.46^{+0.14}_{-1.24}$ \\ \hline
0112970501 & 266.5944 & -28.8720 & $12.10^{+8.12}_{-5.80}$ & $1.61^{+1.01}_{-0.84}$ & $0.54^{+0.12}_{-0.13}$ & $3.23^{+0.82}_{-0.82}$ & $1.13^{a}$ & $4.51^{+0.16}_{-3.86}$ \\ \hline
0762250301 & 266.5952 & -28.8721 & $3.19^{+1.22}_{-0.97}$ & $0.56^{+0.31}_{-0.29}$ & $0.23^{+0.12}_{-0.10}$ & $1.16^{+0.52}_{-0.52}$ & $0.59^{a}$ & $3.85^{+0.24}_{-2.53}$ & CV$^{b}$\\ \hline
0152920101 & 266.6363 & -30.0780 & $4.09^{+1.7}_{-1.10}$ & $0.37^{+0.27}_{-0.24}$ & $0.35^{+0.12}_{-0.10}$ & $4.27^{+1.31}_{-1.31}$ & $2.96^{+1.32}_{-1.32}$ & $9.21^{+0.40}_{-5.64}$\\ \hline
0802410101 & 266.6890 & -28.2638 & $14.20^{+3.61}_{-3.17}$ & $0.24^{+0.35}_{-0.32}$ & $0.74^{+0.08}_{-0.10}$ & $6.89^{+0.72}_{-0.72}$ & $0.68^{+0.05}_{-0.05}$ & $6.45^{+0.24}_{-5.65}$ \\ \hline
0743980401 & 266.7250 & -32.2224 & $3.17^{+2.07}_{-1.54}$ & $0.68^{+0.61}_{-0.53}$ & $0.82^{+0.21}_{-0.21}$ & $2.09^{+0.60}_{-0.60}$ & $1.80^{+0.61}_{-0.60}$ & $2.09^{+0.10}_{-1.91}$ \\ \hline
0112971001 & 266.7262 & -29.2615 & $3.24^{+0.97}_{-0.78}$ & $0.74^{+0.28}_{-0.25}$ & $0.19^{+0.11}_{-0.09}$ & $3.17^{+1.64}_{-1.64}$ & $1.92^{a}$ & $12.9^{+8.0}_{-5.6}$ \\ \hline
0152920101 & 266.7950 & -30.0160 & $4.34^{+3.19}_{-2.15}$ & $0^{+0.35}_{-0.32}$ & $0.27^{+0.10}_{-0.08}$ & $1.14^{+0.44}_{-0.44}$ & $0.37^{a}$ & $3.08^{+0.16}_{-1.97}$ && $2.7^{+4.7}_{-1.4}$ & $2.69\times10^{32}$\\ \hline
0743980401 & 266.8745 & -32.4345 & $1.07^{+0.75}_{-0.52}$ & $0.67^{+0.36}_{-0.32}$ & $0.57^{+0.22}_{-0.25}$ & $2.56^{+1.08}_{-1.08}$ & $1.36^{+1.16}_{-1.16}$ & $3.91^{+0.41}_{-2.43}$ \\ \hline
0694640701 & 267.0173 & -28.2461 & $6.22^{+1.50}_{-1.25}$ & $0.63^{+0.23}_{-0.22}$ & $0.35^{+0.07}_{-0.09}$ & $1.64^{+0.36}_{-0.36}$ & $1.43^{+0.37}_{-0.37}$ & $3.35^{+0.22}_{-1.53}$ && $1.1^{+0.5}_{-0.3}$ & $4.85\times10^{31}$\\ \hline
0802410101 & 267.0213 & -28.4887 & $3.66^{+0.88}_{-0.74}$ & $0.67^{+0.24}_{-0.23}$ & $0.31^{+0.07}_{-0.04}$ & $1.95^{+0.45}_{-0.45}$ & $1.40^{+0.52}_{-0.52}$ & $4.83^{+0.3}_{-2.47}$ && $6.3^{+6.5}_{-3.5}$ & $2.29\times10^{33}$\\ \hline
0694640701 & 267.0704 & -28.1308 & $12.66^{+2.87}_{-2.57}$ & $0.26^{+0.23}_{-0.22}$ & $0.19^{+0.04}_{-0.05}$ & $1.52^{+0.45}_{-0.45}$ & $0.93^{+0.51}_{-0.51}$ & $5.25^{+0.25}_{-2.80}$ & CV$^{b}$ & $2.9^{+2.3}_{-1.4}$ & $5.28\times10^{32}$\\ \hline
0030540101 & 267.1623 & -28.5329 & $2.10^{+1.89}_{-1.28}$ & $-0.43^{+0.48}_{-0.43}$ & $0.30^{+0.15}_{-0.15}$ & $4.01^{+2.05}_{-2.05}$ & $2.93^{+1.75}_{-1.75}$ & $11.1^{+5.0}_{-10.6}$ \\ \hline
0801683201 & 267.1776 & -28.7551 & $1.53^{+0.39}_{-0.33}$ & $0.61^{+0.16}_{-0.15}$ & $0.32^{+0.11}_{-0.09}$ & $2.11^{+0.64}_{-0.64}$ & $1.56^{+0.63}_{-0.63}$ & $5.27^{+0.33}_{-0.9}$ \\ \hline
0205240101 & 267.2002 & -28.2112 & $1.32^{+0.18}_{-0.17}$ & $1.86^{+0.16}_{-0.15}$ & $0.45^{+0.17}_{-0.20}$ & $1.99^{+0.86}_{-0.86}$ & $1.18^{+0.87}_{-0.87}$ & $4.53^{+0.35}_{-0.5}$ && $6.4^{+6.6}_{-2.3}$ & $2.22\times10^{33}$\\ \hline
0801681301 & 267.3237 & -28.5578 & $3.36^{+0.47}_{-0.42}$ & $0.98^{+0.13}_{-0.12}$ & $0.25^{+0.07}_{-0.06}$ & $3.35^{+0.93}_{-0.93}$ & $2.24^{+0.94}_{-0.94}$ & $1.12^{+1.0}_{-2.5}$ & CV$^{b}$ & $5.1^{+4.3}_{-2.4}$ & $3.49\times10^{33}$\\ \hline
0801683401 & 267.4775 & -29.7276 & $1.46^{+0.51}_{-0.42}$ & $0.11^{+0.17}_{-0.16}$ & $0.16^{+0.07}_{-0.09}$ & $2.66^{+1.27}_{-1.27}$ & $0.75^{a}$ & $1.26^{+8}_{+2.1}$ & CV$^{b}$\\ \hline

0844101101 & 267.6210 & -26.6986 & $0.36^{+0.08}_{-0.07}$ & $1.81^{+0.21}_{-0.19}$ & $1.06^{+0.48}_{-0.54}$ & $2.39^{+1.07}_{-1.07}$ & -- & $2.51^{+0.30}_{-0.25}$ && $1.4^{+0.11}_{-0.08}$ & $5.89\times10^{31}$\\ \hline
0886121001 & 267.7746 & -28.0610 & $0.24^{+0.04}_{-0.03}$ & $2.04^{+0.13}_{-0.12}$ & $1.57^{+0.80}_{-0.58}$ & $2.03^{+0.82}_{-0.82}$ & $1.02^{+0.71}_{-0.71}$ & $1.59^{+0.25}_{+0.22}$ && $6.2^{+2.9}_{-2.5}$ & $7.31\times10^{32}$\\ \hline
0801682801 & 268.2557 & -29.2243 & $0.24^{+0.52}_{-0.10}$ & $0.28^{+0.34}_{-0.30}$ & $0.54^{+0.22}_{-0.33}$ & $1.10^{+0.55}_{-0.55}$ & $0.73^{a}$ & $1.74^{+0.19}_{-0.84}$ & CV$^{b}$ & $8.2^{+3.1}_{-2.5}$ & $1.40\times10^{33}$\\ \hline
0932200801 & 268.2677 & -25.5801 & $0.49^{+4.91}_{-0.40}$ & $-0.67^{+0.99}_{-0.60}$ & $0.41^{+0.07}_{-0.21}$ & $1.42^{+0.62}_{-0.62}$ & $0.90^{+0.05}_{-0.05}$ & $2.40^{+0.19}_{-2.10}$ && $3.7^{+2.4}_{-1.2}$ & $3.93\times10^{32}$\\ \hline
0886090401 & 268.4651 & -25.0445 & $3.44^{+4.83}_{-2.83}$ & $-0.41^{+0.82}_{-0.69}$ & $0.49^{+0.18}_{-0.2}$ & $0.29^{+0.09}_{-0.09}$ & $0.91^{a}$ & $4.63^{+0.17}_{-4.10}$ \\ \hline
0679810401 & 268.6678 & -26.2372 & $0.76^{+0.29}_{-0.60}$ & $-0.70^{+0.34}_{-0.39}$ & $0.89^{+0.41}_{-0.32}$ & $6.84^{+3.49}_{-3.49}$ & $5.34^{+4.03}_{-4.03}$ & $8.69^{+0.14}_{-0.14}$ & CV & $8.2^{+2.8}_{-2.5}$ & $6.99\times10^{33}$\\ \hline
0886081301 & 268.7987 & -26.0543 & $0.37^{+0.11}_{-0.09}$ & $1.16^{+0.15}_{-0.16}$ & $0.21^{+0.11}_{-0.11}$ & $0.64^{+0.42}_{-0.42}$ & $0.55^{a}$ & $2.77^{+0.28}_{-0.30}$ & CV$^{b}$ & $4.9^{+3.2}_{-2.4}$ & $7.96\times10^{32}$\\ \hline
0801683101 & 268.8914 & -29.1937 & $0.28^{+0.36}_{-0.23}$ & $-0.31^{+0.36}_{-0.23}$ & $0.38^{+0.10}_{-0.15}$ & $2.16^{+0.81}_{-0.81}$ & $1.76^{+0.82}_{-0.82}$ & $4.57^{+0.56}_{-0.98}$ && $10.1^{+3.1}_{-2.4}$ & $5.58\times10^{33}$\\ \hline
0830190401 & 269.3439 & -25.2988 & $6.31^{+1.85}_{-1.48}$ & $2.45^{+0.71}_{-0.62}$ & $0.93^{+0.36}_{-0.4}$ & $9.92^{+4.34}_{-4.34}$ & -- & $11.5^{+3.6}_{-2.2}$ \\ \hline
0782770201 & 269.4194 & -28.8517 & $4.55^{+2.61}_{-1.91}$ & $0.11^{+0.39}_{-0.31}$ & $0.24^{+0.08}_{-0.11}$ & $1.93^{+0.77}_{-0.77}$ & $1.89^{+0.83}_{-0.83}$ & $5.96^{+0.31}_{-5.70}$ & CV$^{b}$ & $4.5^{+1.8}_{-1.3}$ & $1.44\times10^{33}$\\ \hline
0745060701 & 270.6748 & -20.2881 & $2.06^{+0.20}_{-0.19}$ & $0.98^{+0.08}_{-0.08}$ & $0.04^{+0.01}_{-0.01}$ & $13.6^{+3.3}_{-3.3}$ & $4.67^{+3.41}_{-3.41}$ & $156^{+2}_{-11}$ & HMXB & $8.3^{+4.1}_{-2.4}$ & $1.29\times10^{35}$\\ \hline
0720540401 & 270.9548 & -24.4085 & $5.66^{+3.93}_{-2.71}$ & $3.23^{+1.56}_{-1.26}$ & $1.4^{+0.48}_{-0.55}$ & $1.47^{+0.59}_{-0.59}$ & -- & $1.47^{+0.12}_{-1.4}$ \\ \hline
0720540601 & 270.9673 & -24.3829 & $0.34^{+0.02}_{-0.02}$ & $2.33^{+0.06}_{-0.06}$ & $0.57^{+0.16}_{-0.18}$ & $1.76^{+0.60}_{-0.60}$ & $0.84^{+0.42}_{-0.42}$ & $4.15^{+0.28}_{-0.28}$ & YSO & $1.06^{+0.03}_{-0.03}$ & $5.58\times10^{31}$\\ \hline
0720540501 & 271.0616 & -24.4085 & $1.05^{+0.23}_{-0.19}$ & $2.17^{+0.20}_{-0.18}$ & $0.68^{+0.27}_{-0.29}$ & $2.34^{+0.99}_{-0.99}$ & $1.44^{+0.74}_{-0.74}$ & $4.01^{+0.30}_{-0.66}$ & YSO & $1.1^{+0.2}_{-0.2}$ & $5.81\times10^{31}$\\ \hline
0008820101 & 271.1468 & -24.4587 & $0.29^{+0.03}_{-0.03}$ & $1.71^{+0.08}_{-0.07}$ & $0.47^{+0.22}_{-0.18}$ & $3.74^{+1.43}_{-1.43}$ & $2.07^{+1.22}_{-1.22}$ & $8.21^{+0.64}_{-0.52}$ & YSO & $2.3^{+1.4}_{-0.6}$& $5.20\times10^{32}$\\ \hline
0503170101 & 271.1357 & -21.6690 & $4.57^{+1.14}_{-0.96}$ & $0.31^{+0.22}_{-0.20}$ & $0.31^{+0.07}_{-0.09}$ & $2.71^{+0.67}_{-0.67}$ & $1.59^{+0.75}_{-0.75}$ & $6.46^{+0.27}_{-1.86}$ \\ \hline
0554600301 & 272.2131 & -20.3264 & $1.25^{+0.44}_{-0.35}$ & $0.96^{+0.26}_{-0.23}$ & $0.67^{+0.23}_{-0.21}$ & $1.62^{+0.51}_{-0.51}$ & $1.19^{+0.59}_{-0.59}$ & $2.23^{+0.22}_{-0.57}$ & CV\\ \hline

\hline
\end{tabular}
\tablefoot{Same columns as Table \ref{table:list_source}.}
\end{table*}

\end{appendix}

\end{document}